\documentclass[11pt]{article}

\usepackage[bookmarks=false]{hyperref}
\usepackage{xcolor}
\hypersetup{
    colorlinks,
    linkcolor={red!50!black},
    citecolor={blue!50!black},
    urlcolor={blue!80!black}
}

\usepackage{caption}
\usepackage{amsmath,amssymb}
\usepackage{mathpazo}
\usepackage{tocloft}
\usepackage{siunitx}
\usepackage[english]{babel}
\usepackage{wrapfig}
\usepackage{cite}
\usepackage{pgffor}
\usepackage{pdfpages}
\usepackage{empheq}

\usepackage{color}
\definecolor{blue}{rgb}{0.00,0.00,0.95}

\begin{document}



# Curvature induces active velocity waves in rotating multicellular spheroids

Tom Brandstätter[1,2,*], David B. Brückner[1,3,*], Yu Long Han[4], Ricard Alert[5,6,7,8], Ming Guo[4], Chase P. Broedersz[1,2,+]

[1]Arnold-Sommerfeld-Center for Theoretical Physics, Ludwig-Maximilians-Universität München, Theresienstr. 37, 80333 Munich, Germany
[2]Department of Physics and Astronomy, Vrije Universiteit Amsterdam, 1081 HV Amsterdam, The Netherlands
[3]Institute of Science and Technology Austria, Am Campus 1, 3400 Klosterneuburg, Austria
[4]Department of Mechanical Engineering, Massachusetts Institute of Technology, Cambridge, MA, USA
[5]Lewis-Sigler Institute for Integrative Genomics, Princeton University, Princeton, NJ, USA
[6]Princeton Center for Theoretical Science, Princeton University, Princeton, NJ, USA
[7]Max Planck Institute for the Physics of Complex Systems, Nöthnitzerstr. 38, 01187 Dresden, Germany
[8]Center for Systems Biology Dresden, Pfotenhauerstr. 108, 01307 Dresden, Germany

* These authors contributed equally to this work: Tom Brandstätter, David B. Brückner
+ Corresponding author: c.p.broedersz@vu.nl, Phone: +31 20 59 82953

**The multicellular organization of diverse systems, such as embryos, intestines and tumours, relies on the coordinated migration of cells in 3D curved environments. In these settings, cells establish supracellular patterns of motion, including collective rotation and invasion. While such collective modes are increasingly well understood in 2D flat systems, the consequences of geometrical and topological constraints on collective cell migration in 3D curved tissues are largely unknown. Here, we study 3D collective migration in mammary cell spheroids, which represent a common and conceptually simple curved geometry. We discover that these rotating spheroids exhibit a collective mode of cell migration in the form of a velocity wave propagating along the equator with a wavelength equal to the spheroid perimeter. This wave is accompanied by a pattern of incompressible cellular flow across the spheroid surface featuring topological defects, as dictated by the closed spherical topology. Using a minimal active particle model, we reveal that this collective mode originates from the active flocking behaviour of a cell layer confined to a curved surface. Our results thus identify curvature-induced velocity waves as a generic mode of collective cell migration, which could impact the dynamical organization of 3D curved tissues.**

1

A prominent manifestation of curvature in cellular tissues is the conceptually simple spherical geometry. This geometry arises naturally in a number of *in vivo* systems, including blastocysts[1], egg chambers[2], and tumours[3]. The consequences of the spherical geometry for collective migration can also be studied *in vitro* in systems such as epithelial spheroids[4–6], as well as intestinal[7] and cerebral[8] organoids. Like cells in 2D circular confinements[9–12], migrating cells in 3D spherical confinements often exhibit collective behaviours such as rotational motion[2,13–19]. These and a variety of other collective cell behaviours have been successfully described using active matter theories based on active self-propulsion and alignment interactions between cells[20–22]. Therefore, collectively migrating cells can be placed within a broader class of active matter systems[23], ranging from motile cytoskeletal filaments[24,25] to swarming midges[26] and flocks of birds[27]. Importantly, unlike in 2D, collective cell migration in 3D confinements may be subject to physical constraints imposed by the geometrical curvature and the closed topology of the system[28–31]. Indeed, recent theoretical studies on nematic[32] and polar[33–36] active matter on spherical surfaces illustrate how ordering, collective behaviours, and motion patterns are drastically affected by curvature due to fundamental symmetry and topology principles[37,38]. For example, self-propelled motion occurs along the geodesics of a surface, which are distorted and may intersect due to curvature. In addition, the closed topology of spherical systems prevents states with uniform orientational order, resulting in the formation of topological defects. However, the effects of these geometrical and topological constraints on collective cell migration in inherently curved 3D systems are not understood. These physical constraints may have consequences for the constantly evolving multicellular architecture of the tissue, which controls biological processes ranging from development[1] to cancer progression[3]. Identifying the basic principles of collective migration[39] in 3D curved geometries is thus central for understanding how tissues form, develop shape, and maintain homeostasis.

**Spheroids perform stochastic global rotations**

Here, we study human mammary cell (MCF10A) spheroids as a model system for collective cell migration in spherical geometries. The spheroids are embedded in an alginate and Matrigel-based extracellular matrix. Starting from a single cell, within 5 days of proliferation, a large spheroid is formed comprising of the order of 100 cells with a roughly spherical shape and radii in the range $R \approx 15 - 40$ μm (Fig. 1a,b). These cells are highly motile and migrate in a coordinated fashion, making these spheroids ideal for studying collective modes of 3D cell migration in spherical geometries (Supplementary Movie S1).

To characterize the migratory dynamics, we perform 3D tracking of the trajectories of the cell nuclei throughout each spheroid. These trajectories reveal prominent global rotations of the spheroids, as observed previously[2,13–19], albeit with significant fluctuations (Fig. 1d-f, Supplementary Movie S2). To characterize these stochastic collective rotations, we first calculate the correlation of velocity orientations as a function of intercellular distance $r$: $C(r) = \langle \hat{v}(r_i) \cdot \hat{v}(r_j) | r_{ij} = r \rangle_{i \neq j}$, where $\hat{v}(r_i)$ is the velocity orientation of the $i$-th cell, and

2

we condition on the intercellular distance $r_{ij}$ (Methods). The correlation is positive at short distances, indicating alignment of nearby cell motion (Fig. 1c). At long distances, however, the correlation becomes negative, signifying anti-alignment on opposite sides of the spheroid ($r \approx 2R$). The distance dependence of these correlations agrees with that expected for a global rigid-body rotation (Supplementary Section 3). Such global rotations can be characterized by the spheroid's instantaneous angular speed $\omega$ and rotational order parameter $\Omega$ quantifying alignment of rotational motion (Methods). Both these quantities show that, despite their stochastic cell motion, the spheroids exhibit coherent and persistent global rotations (Fig. 1g-j).

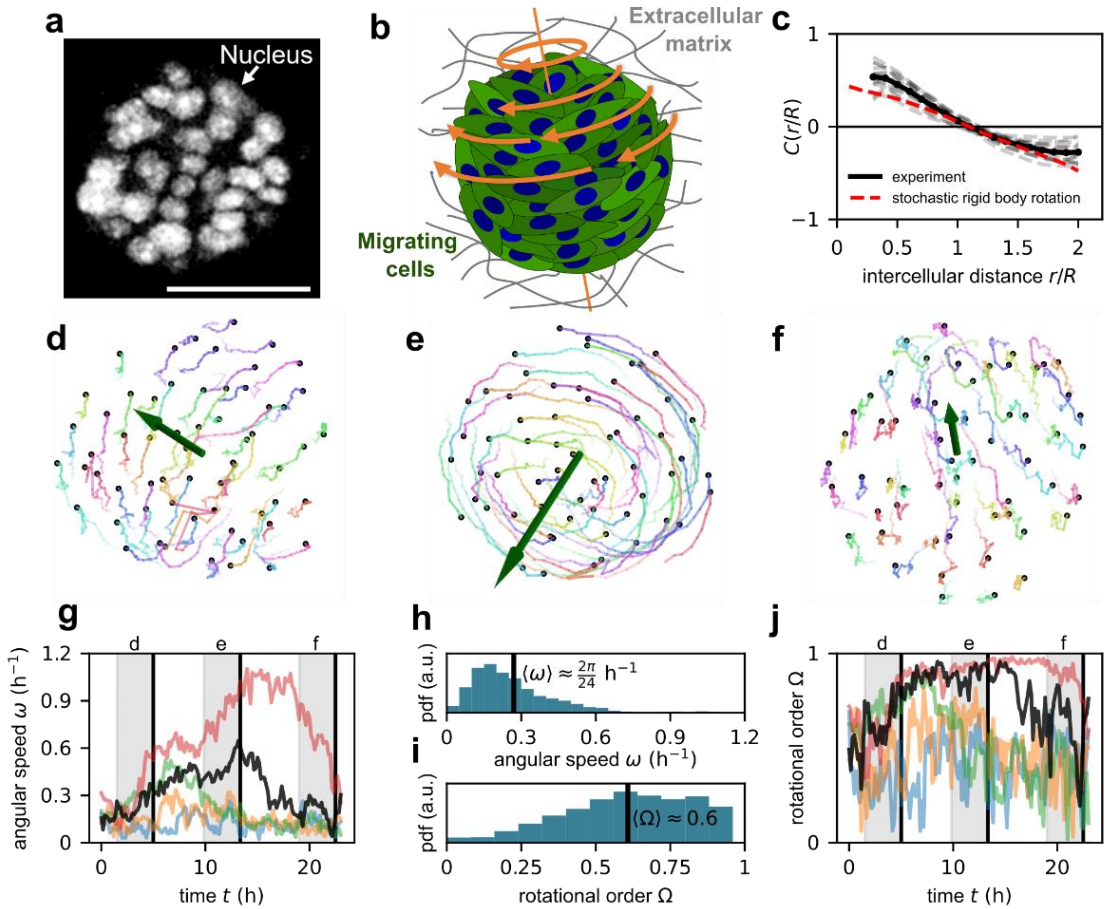

**Figure 1 | Stochastic global rotations of multicellular spheroids. a)** Fluorescence microscopy image of a spheroid with fluorescently labelled (GFP-NLS) nuclei (Methods). Scale bar 50 μm. **b)** Sketch of a rotating spheroid in its extracellular environment. **c)** Spatial correlation function of velocity directions $C(r/R)$, where $r$ is the distance between two cells and $R$ is the spheroid radius (Methods). Experimental curves for individual spheroids (grey) and their average (black) are compared with the correlations of a stochastic rigid-body rotation (red) (Supplementary Section 3). **d)-f)** Time series of cell trajectories in a spheroid. Green vectors show the instantaneous angular velocity vector $\boldsymbol{\omega}$ of the collective rotation. **g)** Time evolution of angular speed $\omega = |\boldsymbol{\omega}|$ in 5 different spheroids. The black curve shows the angular speed of the spheroid shown in d)-f). Grey regions indicate the time period of the trajectories shown in d)-f). **h)** Distribution of angular speed $\omega$. **i)** Distribution of rotational order $\Omega$ (Methods). **j)** Time evolution of rotational order in 5 different spheroids. The black curve shows the rotational order of the spheroid shown in d)-f).

3

**Rotating spheroids exhibit velocity waves**

We next ask whether these spheroids exhibit additional collective migration modes beyond the global rotation. To this end, we study velocity fluctuations $\delta v$ of cells around the global rotation (Methods). These velocity fluctuations are characterized by a non-monotonic correlation function $\tilde{C}(r/R)$ (Methods). At short distances, we find positive correlation, indicating coordinated migration of close neighbours, even beyond the global rotation (Fig. 2a). Pronounced positive correlations appear at large distances up to the spheroid diameter, indicating that cells on opposite sides of the spheroids tend to fluctuate in the same 3D direction. The apparent collapse of $\tilde{C}(r/R)$ for a range of spheroid radii suggests a scale-free correlation structure of the velocity fluctuation field, indicative of a dominant length scale set by the system size. In summary, the correlations of velocity fluctuations reveal an unexpected collective migration mode in these spheroids.

To investigate the nature of this collective mode, we analyse the spatiotemporal structure of the velocity fluctuation field of cells in the equatorial plane (Fig. 2b). Two distinct regions emerge in almost every snapshot: one with cell velocities fluctuating in the direction of the global rotation, and another, on the opposite side of the spheroids, where cell velocities fluctuate against the rotation direction (Fig. 2c). Remarkably, the azimuthal component of velocity fluctuations measured along the equator exhibits an approximately sinusoidal profile, with a wavelength robustly equal to the spheroid perimeter (Fig. 2d, Supplementary Section 2.4). Thus, cells on one side of the spheroid perform faster rotational motion than cells on the opposite side. This fluctuation pattern propagates along the equator (Fig. 2e,f), and hence we identify it as a velocity wave. The speed of wave propagation is approximately equal to that of the global rotation (Fig. 2g, Supplementary Section 2.5). These results thus demonstrate that multicellular spheroids exhibit a velocity wave, which propagates along the equator of the global rotation with a wavelength equal to the spheroid perimeter.

**Velocity waves in incompressible flow**

A key question about the physical nature of this velocity wave is how the cell density responds to velocity gradients in the cellular flow. To investigate this, we first measure the surface density $\rho_s$ of cells in the equatorial region (Supplementary Section 2.8), and find that density variations $\delta\rho_s = \rho_s - \langle\rho_s\rangle$ along the equator are negligible compared to the average density $\langle\rho_s\rangle$. Furthermore, in the frame co-moving with the velocity wave, the surface density exhibits no systematic large-scale deviations from a uniform profile, both in individual time points (Supplementary Fig. 7c) and when averaged over time (Fig. 2h). These results show that the velocity wave in rotating spheroids does not appear to couple to large-scale density variations, as for instance in active matter sound waves[34,40,41].

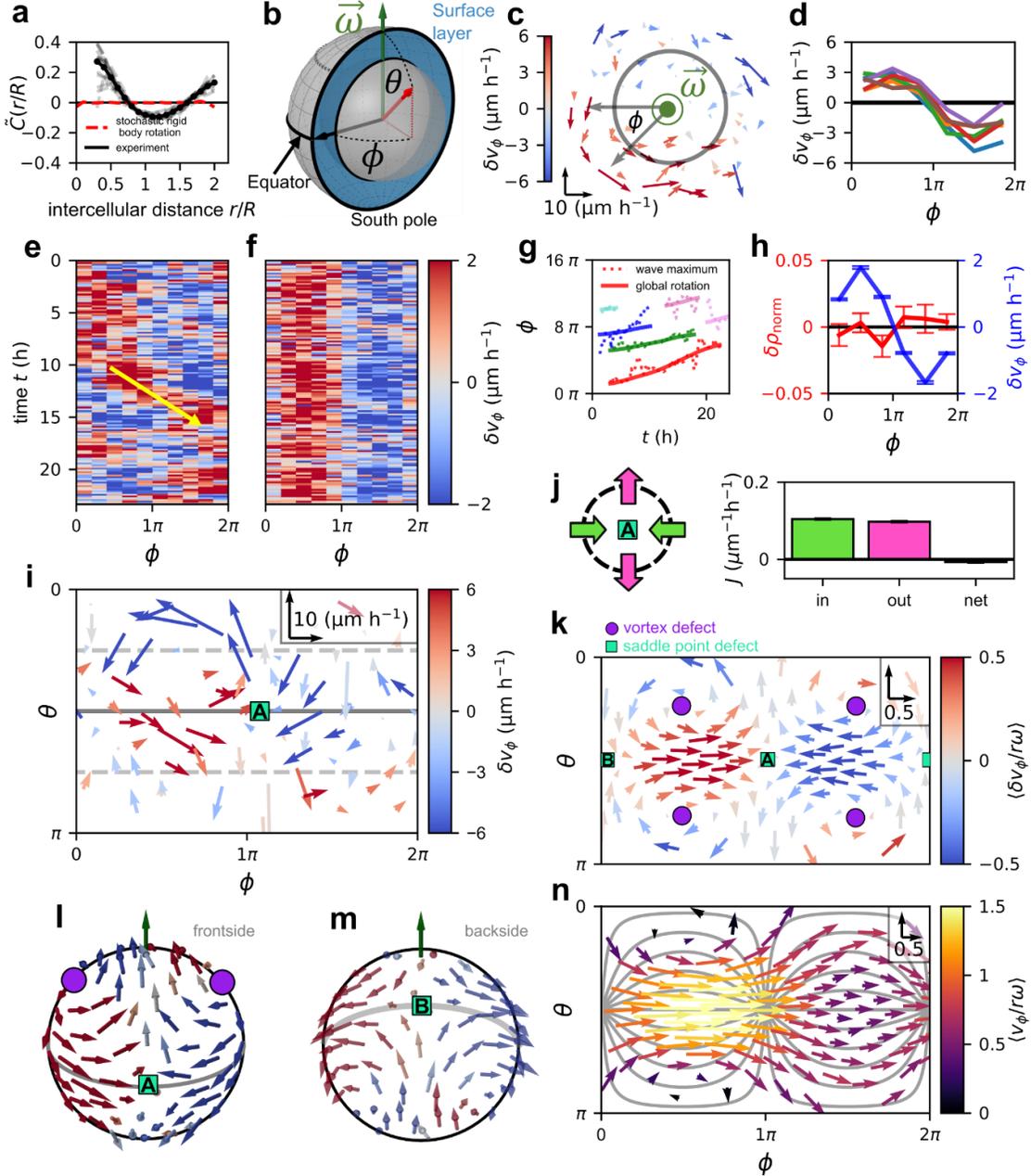

**Figure 2 | Velocity waves in rotating spheroids. a)** Spatial correlation function of velocity fluctuation directions $\tilde{C}(r/R)$ (Methods). The intercellular distance $r$ is scaled by the spheroid radius $R$. As a reference case, we consider a rigid-body rotation with additional uncorrelated Gaussian noise (dashed red line) (Supplementary Section 3). **b)** Schematic of the spherical coordinate system. The blue region indicates the cross section of the surface layer of the spheroids. The core region is indicated by the gray inner sphere. We choose the threshold between core region and surface layer at $r = 0.6R$. **c)** Planar projection of the velocity fluctuation field $\delta v$ in a slab centered around the equatorial plane. The color code indicates $\delta v_\phi$, the $\phi$-component of $\delta v$, and the arrow scale is shown in the lower left corner. The grey circle indicates the threshold between the core and the surface layer. **d)** Azimuthal profile of velocity fluctuations along the equator, $\delta v_\phi(\phi)$, in the frame of reference co-moving with the wave. Each curve was obtained by averaging over a time interval of $\Delta t = 100$ min (10 frames) around a certain time point in a different spheroid. **e,f)** Kymograph of the azimuthal component of velocity fluctuations, $\delta v_\phi(\phi, t)$, along the equator of one spheroid, both in the COM frame (e) and in the frame co-moving with the wave (f) (Methods) **g)** Wave propagation in different spheroids. The solid line shows the integrated angular speed giving the total path length of the

5

global rotation. Dashed lines indicate the trajectory of the propagating velocity wave maximum. We consider trajectories with a minimum length of 15 time points (Supplementary Section 2.5). **h)** Normalized surface density profile $\delta\rho_{norm} = (\rho_s - \langle\rho_s\rangle)/\langle\rho_s\rangle$ along the equator averaged over all spheroids and time points (red). Velocity fluctuation profile in the equator averaged over all spheroids and time points (blue). Error bars represent the standard error of the mean (s.e.m). **i)** Snapshot of the tangential velocity fluctuation field, $\delta v^t$, in the surface layer of an individual spheroid represented in spherical coordinates $\phi$ and $\theta$ (Methods). Inset: arrow scale. **j)** Flux analysis around the saddle point ahead of the velocity wave maximum. Bar plot shows the absolute values of the average influx and outflux. Black bar indicates the netflux. **k)** Tangential components of the average velocity fluctuation field $\langle\delta v/r\omega\rangle$ (Methods) in spherical coordinates including its six topological defects. **n)** Tangential components of the average tangential total velocity field $\langle v/r\omega\rangle$ in spherical coordinates. We show the geodesics of the sphere, which are great circles. As in k), velocity fields are averaged over all spheroids and timepoints (Supplementary Section 2.7). **l,m)** Frontside (l) and backside (m) of the average velocity fluctuation field. Both are shown from the same perspective, with the backside field shown by looking through the sphere.

We next ask how the cellular flow in the spheroid balances the velocity gradients in the equatorial velocity wave while avoiding equatorial density fluctuations. Importantly, there is no significant cellular exchange between the surface and the core, as shown by the absence of significant radial cellular flows (Supplementary Section 2.6). Therefore, we analyse the motion of cells in the surface layer (Fig. 2b), which exhibits the largest velocity fluctuations that are dominated by components tangential to the spheroid surface (Fig. 2c, Supplementary Fig.4 a,c-i). Snapshots of these tangential velocity fluctuations $\delta v^t$ (Methods) in spherical coordinates reveal that cell motion features significant polar components near the high velocity gradients of the equatorial velocity wave (Fig. 2i). This pattern is organized by incompressible tangential flow across the whole spheroid surface. We show this by measuring the tangential cell flux $\rho_s \delta v^t$ in the surface layer (Methods). While there are significant cellular fluxes, their divergences are negligible (Supplementary Section 2.9), even around the extrema of the velocity gradients (Fig. 2j), as required for incompressible flow.

The incompressible nature of the cellular flow is further illustrated by the average velocity fluctuation field in the frame co-moving with the wave (Fig. 2f,k-m, Methods). The large-scale pattern in this field features a total of six topological defects, at which the velocity direction is undefined. This pattern is robust: it is not only visible in the population average, but also appears instantaneously in individual spheroids (Supplementary Section 2.7.2). The four vortex defects carry a topological charge of $q_v = +1$, while the charge of the two saddle point defects is $q_s = -1$. The topological charges add up to $q_{\text{total}} = 4q_v + 2q_s = 2$, as dictated by the Poincaré-Hopf theorem for continuous vector fields on a sphere. Thus, the collective supracellular pattern of cell motion on the spheroid surface obeys constraints set by the system's closed spherical topology.

In the average total velocity field of the spheroids, the defect structure manifests as a diverging and converging pattern of motion around the azimuthal position ($\phi \approx 3\pi/2$) of the velocity wave minimum (Fig. 2n). This motion pattern appears to be structured along the geodesics of the spherical confinement. Altogether, the patterns in the velocity and the velocity fluctuation fields show how cells ahead of the velocity wave maximum divert towards the poles without increasing local density or moving radially. Cells at the poles flow into the equatorial region

6

behind the wave maximum where cells accelerate again, thereby generating a global supracellular pattern of incompressible cell motion shaped by the spherical geometry.

**Active particles on spheres exhibit velocity waves**

The topological defect structure (Fig. 2k) and the role of geodesics in directing cell motion (Fig. 2n) suggest that topological and geometric features owing to the spherical confinement (Fig. 3a,b) constrain the observed collective migration modes in spheroids. To theoretically elucidate the physical implications of the spherical geometry for the collective dynamics in the surface layer of rotating spheroids, we employ a minimal biophysical model for a layer of active particles (cells) constrained to a sphere[33]. This model features common aspects of collective cell migration[21], including self-propulsion along a polarity vector, which aligns with the polarity of neighbouring cells with strength $\beta$ and is subject to dynamical noise with amplitude $\sigma$ (Methods). Here, we focus on the largely unexplored limit of an incompressible confluent spherical cell layer by imposing strong repulsion interactions between cells.

We vary the two key parameters $\beta$ and $\sigma$ in this model to construct a phase diagram for active incompressible multicellular motion on the sphere (Fig. 3c). For weak alignment, we find a disordered phase, exhibiting neither polarity alignment nor global rotations. When alignment interactions are strong, the system organizes into an ordered but *quiescent* regime with a polarity field containing two aster defects (Fig. 3f), similar to a ground state of an aligned vector field on a sphere (Fig. 3b). In contrast, at intermediate alignment strengths, we find a regime with persistent global rotations (Fig. 3d). The collective dynamics in this phase are remarkably similar to our experimental observations (Fig. 3e). In fact, we identify parameter values for which the model quantitatively reproduces the experimental distribution of rotational order and the correlation function of velocity fluctuation directions (Fig. 3j,k).

Within the collective rotation regime, the model robustly predicts the emergence of velocity waves (Fig. 3h,i, Supplementary Section 5.1). As in the experiments, the wave predicted by the model consists of a single-wavelength velocity modulation along the equator, accompanied by four vortices (Fig. 3l,m). The two vortex defects in the polarity field are displaced away from the poles, and therefore drive a state of global rotation with polarized cell flows over the poles. This indicates that cells in the model actively migrate towards and over the poles (Fig. 3g, Supplementary Section 5.4). Furthermore, the model predicts that the wave propagates approximately with the same angular speed as the global rotation: $\omega_{\text{wave}} \approx \omega$, consistent with our experiments (Fig. 2g, Supplementary Section 2.5). Finally, we find that the emergence of the velocity wave does not depend sensitively on the specific form of the alignment interactions between the self-propelled particles (Supplementary Section 5.3). Our results thus suggest that the velocity wave and the accompanying vortices in the rotating spheroids are a generic collective mode of active particles confined to a spherical surface.

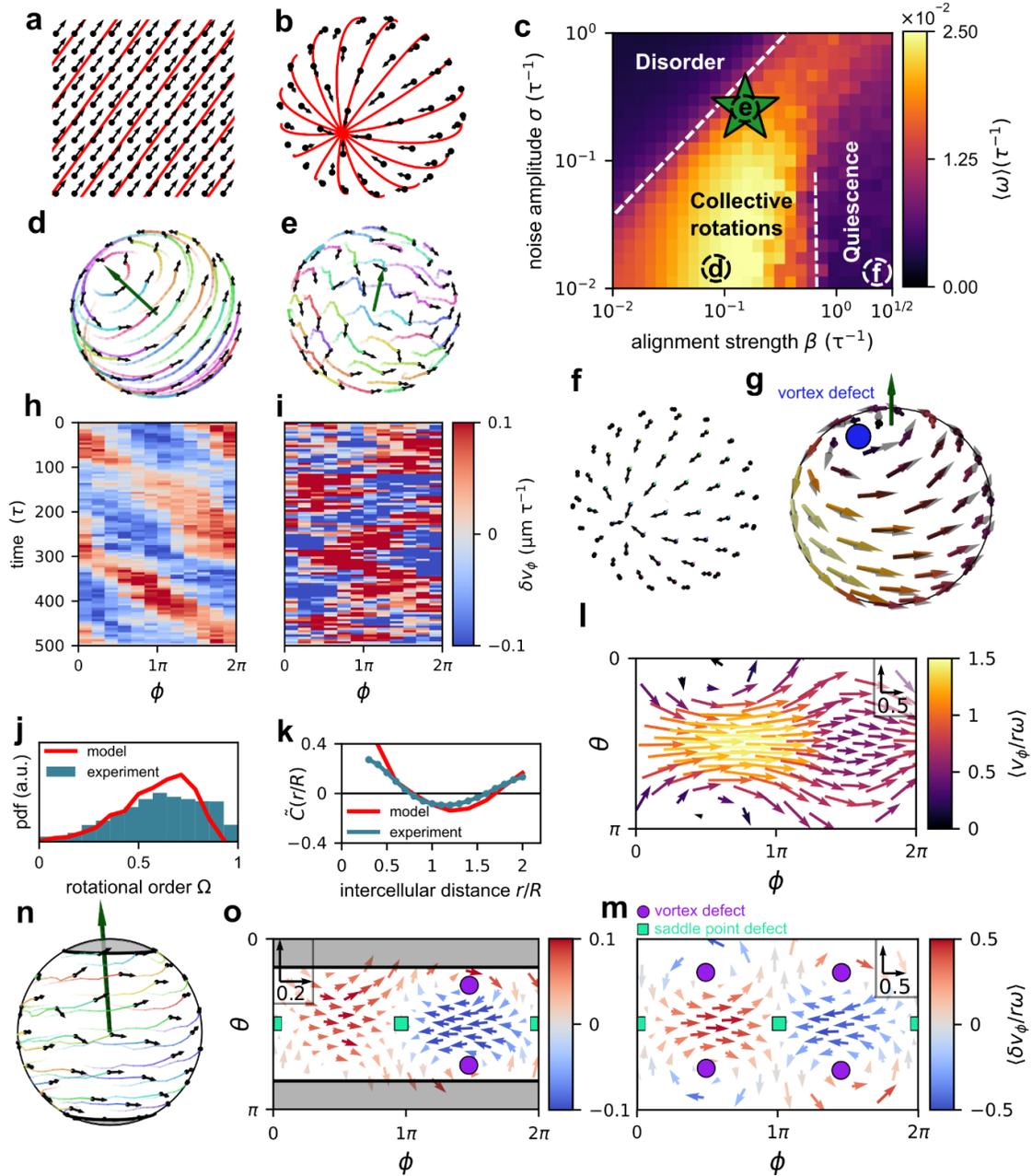

**Figure 3 | Velocity waves in a model for active particles on a sphere. a),b)** Ground states of an aligning vector field on a flat surface (a) and on a sphere (b). Both are found by implementing alignment interactions between vectors at fixed positions. Red lines show the geodesics of the different geometries. On a spherical surface no global order is possible and geodesics intersect. **c)** Phase diagram of active aligning particles constrained to a sphere constructed from the average angular speed ⟨ω⟩. The timescale $\tau$ is defined through the speed of self-propulsion in our model (Supplementary Section 4). The green star indicates the parameter value ($\beta = 0.079$, $\sigma = 0.31$), where the model quantitatively reproduces the rotational dynamics measured experimentally. White dashed lines suggest apparent phase boundaries. **d)-f)** Snapshots of predicted states marked in the phase diagram. Gray vectors are particle polarities, trajectories are plotted as colored lines fading over time (Supplementary Movie 3, 4). The green vector shows the axis of rotation of the layer of active particles. **g)** Average velocity field for the experimental parameter combination (green star in panel c) plotted on a sphere together with the average polarity

8

field. **h,i)** Representative kymograph $\delta v_\phi(\phi, t)$ of a velocity wave predicted by our model in the regime of deterministic collective rotation ($\beta = 0.2$, $\sigma = 0.012$) (h) and for the experimental parameter values (i). **j)** Distribution of rotational order $\Omega$ in the model (red) and the experiments (black). **k)** Spatial correlation of velocity fluctuation directions $\tilde{C}(r/R)$ in the model (red) and the experiments (blue). **l)** Tangential components of the average tangential total velocity field $\langle v/r\omega \rangle$ in spherical coordinates for the experimental parameter combination. **m)** Tangential components of the average velocity fluctuation field $\langle \delta v/r\omega \rangle$ in spherical coordinates including its six topological defects. We show the result for the low-noise regime in Supplementary Fig. 10e. **n)** Snapshot of a simulation on a sphere from which we removed two opposing spherical caps (grey). **o)** Tangential components of average velocity fluctuation field in the simulation without two opposing caps (grey regions) represented in spherical coordinates. Instead of four vortex defects, this system exhibits only two vortex defects, leading to a total topological charge of zero.

**Curvature induces active velocity waves**

Having demonstrated that active velocity waves occur in an incompressible layer of active particles constrained to a sphere, we next investigate which properties of the sphere are responsible for their formation. To this end, we use our simulations to disentangle the role of the two key properties of the spherical geometry: positive Gaussian curvature and a closed topology. By removing two opposing spherical caps from the sphere, we retain positive curvature but change the closed to an open topology (Fig. 3n). Remarkably however, the velocity wave still emerges, indicating its robustness to topological changes of the confinement (Fig. 3o, Supplementary Section 5.5.1). Importantly, for this geometry, the total topological defect charge $q_{\text{total}}$ must be equal to zero, as we indeed observe in our simulations (Fig. 3o). In contrast, on cylinders, which have zero Gaussian curvature, we do not observe a single-wavelength velocity wave propagating in the direction of the global rotation (Supplementary Section 5.5.2). Taken together, these results demonstrate that the curvature and topology of multicellular spheroids play distinct roles in determining the collective dynamics: While the topological defect structure of the global flow pattern is constrained by the spherical topology, the active velocity waves themselves are induced by Gaussian curvature.

**Discussion**

In conclusion, we discover a collective mode of cell migration manifesting as an active velocity wave modulating the global rotations of mammary cell spheroids. Previously reported wave-like phenomena in cell migration and active matter include mechanical[42] and mechanochemical[43] waves in spreading epithelial monolayers, as well as sound waves in active matter flocks that have been predicted theoretically on both flat[40,41,44] and curved[34] surfaces. However, all those waves occur in compressible systems, and they are coupled to significant density modulations. In contrast, the velocity waves we discover here both experimentally and theoretically exhibit no appreciable density modulations. These incompressible velocity waves are embedded in a globally-structured flow field, resulting in a supracellular pattern of motion over the entire spheroid surface. We demonstrate theoretically that this pattern emerges as a response of an interacting active matter system to the constraints imposed by the spherical geometry. Conceptually, the active velocity wave is thus induced by the presence of Gaussian curvature, which enforces ordered cell motion along the converging geodesics of the curved

tissue. In addition, the closed topology of the system dictates the presence of topological defects, manifesting as four vortex defects and two saddle point defects in the flow field, similar to recent observations in fly embryos[45]. A key feature of the incompressible flow field is the presence of saddle-point defects where equatorial cell fluxes are balanced by fluxes towards the poles of the spheroid. Cell fluxes near topological defects were observed to drive layer formation in bacterial colonies[46], and they are related to mechanical stresses and sites of cell death in epithelial monolayers[47], suggesting that the reported active velocity wave could play a role in morphological changes in curved tissues. Taken together, our work reveals curvature-induced collective active-matter modes in spherical tissue architectures, with potential implications for cell migration in curved tissues such as embryos and tumours.

# Methods

**Mammary cell spheroid culture and immunofluorescence staining**

MCF10A cells were purchased from ATCC and cultured in a DMEM/F12 medium (Invitrogen, 11965-118) supplemented with 20 ng/ml epidermal growth factor (Peprotech, AF-100-15), 5% horse serum (Invitrogen, 16050-122), 100 ng/ml cholera toxin (Sigma, C-8052),0.5 μg/ml hydrocortisone (Sigma, H-0888), 10 μg/ml insulin (Sigma, I-1882) and 1% penicillin and streptomycin (Thermo Fisher, 15140122). To prepare the mammary spheroid, MCF10A cells were harvested using a 0.05% trypsin solution (Thermo Fisher, 25300054) and seeded in a collagen network (3.6 mg/ml, FibriCol®, Catalog #5133) supplied with Matrigel (2.0 mg/ml, Corning, 354234). The initial cell density is low ($10^4$ /ml) to avoid interaction between spheroids. To visualize the cell nuclei, the MCF10A cells were transfected with GFP-NLS using lentivirus (Essen Bioscience, 4475). The three-dimensional motion of the cells within the mammary spheroid was recorded using confocal microscopy (Leica, TCL SP8). The temperature, CO2 level, and humidity were well-controlled using a culture box during the imaging process. The cell nuclei were further tracked from the video with Trackmate (v5.2.0, Fiji)[48].

**Analysis of trajectories**

All 3D nucleus positions are transformed into the centre of mass (COM) frame of the spheroids. Translational motion of the spheroids is small compared to the spheroid radii (Supplementary Section 2.1). We write $r_i$ for the positions of the cell nuclei in the COM frame. Cell velocities are estimated by numerical derivates of the cell nucleus trajectories, $v_i(t) = (r_i(t + \Delta t) - r_i(t))/\Delta t$, where $\Delta t = 10$ min is the observation interval. To characterize rotations, we infer the rotation matrix $\mathcal{R}(t)$ of each spheroid at each point in time, which describes the average rotation of all cells in the spheroid between subsequent frames. We infer this matrix by simultaneously minimizing the cost functions $E_k^2 = \sum_{i=0}^{N}\left(r^k{}_i(t + \Delta t) - \tilde{r}^k{}_i(t)\right)^2$, where $\tilde{r}_i(t) = \mathcal{R}(t)r_i(t)$ and $k = 1,2,3$ indicates the three spatial dimensions[49]. The angular velocity, $\boldsymbol{\omega}(t)$, is then defined by $\mathcal{R}(t)\boldsymbol{\omega}(t) = \boldsymbol{\omega}(t)$, the angular speed is $\omega(t) = |\boldsymbol{\omega}(t)|$, and the axis of rotation is $\hat{\boldsymbol{\omega}}(t) = \boldsymbol{\omega}(t)/\omega(t)$. Furthermore, we compute the rotational order parameter by[26]:

$$\Omega(t) = \frac{1}{N}\left|\sum_i \frac{r^\perp{}_i(t) \times v_i(t)}{|r^\perp{}_i(t) \times v_i(t)|} \cdot \hat{\boldsymbol{\omega}}(t)\right|$$

where $r^\perp{}_i = r_i - (r_i \cdot \hat{\boldsymbol{\omega}})r_i$ is the orthogonal component of the positions of the cells with respect to the axis of rotation $\hat{\boldsymbol{\omega}}(t)$ and $N$ is the number of cells. We determine the velocity fluctuations of cells around global rotations using $\delta v_i = (r_i(t + \Delta t) - \mathcal{R}(t)r_i(t))/\Delta t$. The cross correlation function of the velocity- and velocity fluctuation directions are given by[26]:

$$C(r) = \frac{\sum_{i \neq j}^{N} \delta(|r_{ij}|-r)\hat{v}_i\hat{v}_j}{\sum_{i \neq j}^{N} \delta(|r_{ij}|-r)}, \tilde{C}(r) = \frac{\sum_{i \neq j}^{N} \delta(|r_{ij}|-r)\delta\hat{v}_i\delta\hat{v}_j}{\sum_{i \neq j}^{N} \delta(|r_{ij}|-r)}$$

where $\hat{v}_i = v_i/|v_i|$ and $\delta\hat{v}_i = \delta v_i/|\delta v_i|$ and $r_{ij} = r_i - r_j$. We approximate the Dirac-delta function $\delta(|r_{ij}| - r)$ by discrete binning of the intercellular distances. For more details see Supplementary Section 2.3.1.

**Analysis of vector fields**

We consider the set of all velocities $\{v_i(t)\}$ at their respective positions $r_i$ as the velocity field $v(r, t)$ of the spheroids at time $t$. Similarly, we consider $\{\delta v_i(t)\}$ as the velocity fluctuation field $\delta v(r, t)$. We rotate the velocity- and the velocity fluctuation field such that the instantaneous axis of rotation describing the rotational

11

motion patterns is aligned to the z-axis of a new coordinate frame $O'$. We can formulate this as a linear transformation $\mathcal{T}(t)$ equivalent to a rotation of the vector fields: $\mathbf{r}' = \mathcal{T}(t)\mathbf{r}$, $\mathbf{v}' = \mathcal{T}(t)\mathbf{v}$, $\delta\mathbf{v}' = \mathcal{T}(t)\delta\mathbf{v}$, where $\mathcal{T}(t)$ is constructed from the axis of rotation (Supplementary Section 2.3.2). Henceforth, for simplicity we omit the primes in the coordinates of the rotated frame. We represent the velocity- and the velocity fluctuation field by the usual spherical coordinates $(\phi_i, \theta_i, r_i)$ for the positions of the $i$-th cell. The angle $\phi \in [0, 2\pi]$ parameterizes the azimuthal angle and the angle $\theta \in [0, \pi]$ parameterizes the polar angle. We define tangential velocity fluctuations of the $i$-th cell as $\delta\mathbf{v}^{\mathrm{t}}(\mathbf{r}_i, t) = \delta\mathbf{v}(\mathbf{r}_i, t) - [\hat{e}_{r_i} \cdot \delta\mathbf{v}(\mathbf{r}_i, t)]\,\hat{e}_{r_i} = [\hat{e}_{\phi_i} \cdot \delta\mathbf{v}(\mathbf{r}_i, t)]\hat{e}_{\phi_i} + [\hat{e}_{\theta_i} \cdot \delta\mathbf{v}(\mathbf{r}_i, t)]\,\hat{e}_{\theta_i}$ and similarly the tangential velocity field $\mathbf{v}^{\mathrm{t}}(\mathbf{r}_i, t) = [\hat{e}_{\phi_i} \cdot \mathbf{v}(\mathbf{r}_i, t)]\hat{e}_{\phi_i} + [\hat{e}_{\theta_i} \cdot \mathbf{v}(\mathbf{r}_i, t)]\,\hat{e}_{\theta_i}$. We show the $\phi$- and the $\theta$- components of the velocity fluctuations as a vector field in Fig. 2i. When we employ this representation, we restrict the velocity fields to the surface layer defined by $|\mathbf{r}| > r_{\mathrm{thresh}} = 0.8R$ where $R$ is the radius of the spheroids. The spheroid radius $R$ is determined by creating a convex hull around the positions of the cells and averaging the distance of the outer most cells to the COM (Supplementary Section 2.2). The pattern we observe in this representation is not sensitive to the choice of $r_{\mathrm{thresh}}$.

**Kymographs of velocity waves**

We find the kymograph of the equatorial azimuthal fluctuation profile in the surface layer of the spheroids by defining $N_b$ bins along the azimuthal $\phi$-direction with width $d\phi$. We furthermore focus on the equatorial region defined by $\theta \in [\pi/2 - d\theta, \pi/2 + d\theta]$. We choose $d\theta = \pi/4$, $N_{\mathrm{b}} = 10$ and $d\phi = \pi/5$, and focus on the surface layer defined by $|\mathbf{r}| > r_{\mathrm{thresh}} = 0.6R$. The $\phi$-component of the velocity fluctuations at time t: $\hat{e}_{\phi_i} \delta\mathbf{v}(\mathbf{r}_i, t)$ is averaged inside these bins to obtain the kymograph $\delta v_\phi(\phi, t)$. The resulting kymograph is insensitive to the choice of $r_{\mathrm{thresh}}$ (Supplementary Fig. 2g).

**Averaging of vector fields**

The average velocity- and velocity fluctuation field in the frame of reference of the propagating velocity wave are found by first rotating $\mathbf{v}(\mathbf{r}, t)$ and $\delta\mathbf{v}(\mathbf{r}, t)$ around the axis of rotation $\widehat{\boldsymbol{\omega}}(t)$ so that all wave maxima are aligned on the azimuthal angle $\phi = \pi/2$. The angle by which we rotate is found through fitting a single-wavelength sinusoidal profile to the individual equatorial fluctuation profiles shown in the kymograph (Fig. 2e). Then the velocity and velocity fluctuation field in the surface layer are scaled down for each individual cell by $\omega(t)|\mathbf{r}_i|$ to make them dimensionless. This yields $\mathbf{v}(\mathbf{r}_i, t)/\omega(t)|\mathbf{r}_i|$ and $\delta\mathbf{v}(\mathbf{r}_i, t)/\omega(t)|\mathbf{r}_i|$. Finally, the surface layer of the spheroids is covered in $N_{\mathrm{bins}} = 120$ uniformly distributed spherical bins. The size of each bin is defined by an angle $d\Omega = 0.2$ between the position vector of each bin and the position vector of a cell. This results in sufficient covering of the surface layer with bins. All rescaled velocities and velocity fluctuations inside one bin are averaged over time and/or different experimental realizations. Finally, we find the tangential components of the average vector fields by employing the same procedure as for the individual snapshots. For more details see Supplementary Section 2.7.

**Flux analysis**

We find the tangential cell flux $\rho_s \delta\mathbf{v}^{\mathrm{t}}$ where $\rho_s$ is the approximated surface density of cells in the surface layer. We then construct a circle around the position $\mathbf{r}_0$ to probe the incompressibility of the cell flow at this position. The size of the circle is characterized by $\Omega$ which is the angle between the positions of the cells $\mathbf{r}_i$ and $\mathbf{r}_0$. We find the cell flux through this circle by considering all cells whose angle is within $[\Omega - d\Omega, \Omega + d\Omega]$. We project the cell flux on the normal vectors to this circle and sum all positive values and all negative values to find the out- and the influx. The sum of all projections is equal to the net cell flux. For more details see Supplementary Section 2.9.

**Active particles on curved surfaces**

Based on previous work[50–53], we model cells as active particles. We describe the motility of cells by self-propulsion with speed $v_0$ in the direction of an internal polarization defined by the polarity vector $\boldsymbol{p}$. We impose an alignment interaction with strength $\beta$ between polarities of neighbouring cells and polarities are subject to Gaussian white noise with amplitude $\sigma$. In addition, cells repel each other by a soft repulsion potential with stiffness $\varepsilon$. On a flat surface the model is written as:

$$\boldsymbol{v}_i = v_0 \boldsymbol{p}_i + \boldsymbol{F}_{\text{rep}}, \quad \boldsymbol{F}_{\text{rep}} = -\varepsilon(2\lambda - r_{ij})\frac{r_{ij}}{r_{ij}}, \quad \boldsymbol{p}_i = (\cos(\phi_i), \sin(\phi_i))^T, \quad \frac{d\phi_i}{dt} = -\beta \sum_j \sin(\phi_i - \phi_j) + \eta_i(t)$$

$$\langle \eta_i(t)\eta_j(t') \rangle = \sigma^2 \delta_{ij} \delta(t - t')$$

where $\lambda$ is the radius of the particles and $\sum_j$ is the sum over all neighbours of the *i*-th particle within by a radius of interaction $r_{\text{inter}} = 2.5\lambda$. To model the dynamics of the surface layer of the spheroids, we study this model on a surface of a sphere[33]. Importantly, we consider the largely unexplored incompressible limit of this model by setting the repulsion stiffness $\varepsilon$ to a sufficiently high value such that the layer of cells remains approximately confluent without jamming. We set the time scale of our simulation to $\tau = l/v_0$, where $l = 1\mu m$ is the length scale of our simulation. We set the average surface density of cells on a sphere in our simulation to the average surface density of the cells in the surface layer of the spheroids. Together with the number of cells, this also sets the size of the sphere and the radius of cells. We numerically solve this model on a sphere by an algorithm based on the Euler-Maruyama method[33] (Supplementary Section 4). Furthermore, we vary the underlying geometry from a sphere to a truncated sphere and a cylinder, which are both unexplored for this model. For the truncated sphere, we introduce soft boundaries at two given latitudes. These boundaries keep the cells from moving over the poles. We make sure that the effectively excluded regions are large enough so that there are no alignment interactions between cells on opposite sides of the excluded regions (Supplementary Section 5.5.1). For the cylinder, we use cylindrical coordinates to constrain the dynamics. This asymmetric geometry requires adjustment of how alignment interactions are implemented. For more details refer to Supplementary Section 5.5.2.

**Data & code availability**

Experimental and simulation data as well as python code is available from the corresponding author upon request.

## Author Contributions

Y.L.H. and M.G. designed experiments; Y.L.H. performed experiments; T.B. and D.B.B. analyzed data; T.B. developed and performed simulations; T.B., D.B.B., R.A. and C.P.B. interpreted the results and wrote the manuscript. M.G. and C.P.B. conceived this study and C.P.B led the research. All authors edited and approved the manuscript.

## Competing interests

The authors declare no competing interests.

## Materials & Correspondence

Supplementary information is available for this paper. Correspondence and request for materials should be addressed to C.P.B. (c.p.broedersz@vu.nl, Phone: +31 20 59 82953).



\newpage

\centerline{\bf{\LARGE{Supplementary Information:}}}\vspace{0.1 cm}
\centerline{\bf{\large{Curvature induces active velocity waves in rotating multicellular spheroids}}}
\vspace{0.5cm}
\centerline{\normalsize{Tom Brandstätter*, David B. Brückner*, Yu Long Han, Ricard Alert, Ming Guo, Chase P. Broedersz$^\dagger$}}
\vspace{0.4 cm}

* These authors contributed equally to this work: Tom Brandstätter, David B. Brückner\\
$\dagger$ Corresponding author: c.p.broedersz@vu.nl, Phone: +31 20 59 82953

%------------------------------------------------------------

\tableofcontents
\newpage
%------------------------------------------------------------
\section{Supplementary movie descriptions}

\textbf{Supplementary Movie 1}\\
Time lapse of fluorescence microscopy images of a cell spheroid. Cell nuclei are fluorescently labelled and appear in white in the video. Scale bar: $50 \   \ \si{\micro\meter}$.

\textbf{Supplementary Movie 2}\\
Collective dynamics of all cell spheroids. Trajectories are shown as gray lines. Cell spheroids are ordered according to their radius.

\textbf{Supplementary Movie 3}\\
Collective dynamics of our model as predicted in the experimental parameter regime. Green vector shows the instantaneous angular velocity of the cells. Colored lines show the trajectories of cells. 

\textbf{Supplementary Movie 4}\\
Collective dynamics of our model as predicted in the low-noise parameter regime. Green vector shows the instantaneous angular velocity of the cells. Colored lines show the trajectories of cells. 

\textbf{Supplementary Movie 5}\\
Collective dynamics of our model as predicted on a sphere where we removed two opposing caps. 

\textbf{Supplementary Movie 6}\\
Collective dynamics of our model as predicted on a cylinder. 

\newpage

%------------------------------------------------------------
\section{Supplementary experimental analysis}

\subsection{Center of mass motion of cell spheroids}

In the main text, we analyze the rotational mode of the system in the centre of mass (COM) frame. Here we characterize the motion of the COM. To this end, we find the trajectory of the COM in the lab frame by: 
\begin{equation}
	\vec{r}_{\text{COM}}(t) = \frac{1}{N_{\text{cells}}}\sum_{i=1}^{N_{\text{cells}}}\vec{r}_{i}(t)
\end{equation}
Here, $\{\vec{r}_{i}(t)\}$ is the set of all (lab frame) positions of the cells in one spheroid and $N_{\text{cells}}$ denotes the number of cells. Furthermore, we quantify the velocity of the COM by 
\begin{equation}
	\vec{v}_{\text{COM}}(t) = (\vec{r}_{\text{COM}}(t+\Delta t)-\vec{r}_{\text{COM}}(t))/\Delta t
\end{equation}
where $\Delta t = 10 \text{min}$ is the observation interval. We note that the instantaneous displacement of the COM $|\vec{v}_{\text{COM}}(t)|\Delta t$ is small with respect to the spheroid radius $R$ (Fig. \ref{fig:spheroid_overview}a). We explain how we compute $R$ in section \ref{sec:description}. To further characterize the COM motion of spheroids, we compute the Mean squared displacement (MSD) of the COM position of the spheroids, which we here define as $MSD(\tau) = \langle|\vec{r}_{\text{COM}}(t) - \vec{r}_{\text{COM}}(t+\tau)|^2\rangle_t$. Here $\langle...\rangle_t$ denotes an average over all time points. We find that COM motion is subdiffusive. Importantly, at the end of the observation period, the average root mean squared displacement of the COMs is around $\sqrt{10} \ \mu m$ (Fig. \ref{fig:spheroid_overview}b), which is small compared to the spheroid radii (Fig. \ref{fig:spheroid_overview}d). Furthermore, we consider the auto-correlation function of the velocity of the COM:
\begin{equation}
	\chi(\tau) = \langle\vec{v}_{\text{COM}}(t)\vec{v}_{\text{COM}}(t-\tau)\rangle_{t}
\end{equation}
This quantity does not exhibit a characteristic time scale, which would be related to directed COM motion of collectively translating spheroids (Fig. \ref{fig:spheroid_overview}c). Altogether, these results indicate the absence of persistent COM motion, meaning that cell spheroids remain approximately fixed at their lab positions. Note that throughout the main text and the supplementary information, all data analysis is conducted in the COM frame. Therefore, henceforth, we write the positions of the cells in the COM frame as $\vec{r}_{i}(t)$.

\subsection{Size, shape and number of cells of spheroids}\label{sec:description}

In this subsection, we give an overview of the $N_{s} = 16$ spheroids and characterize their shape as well as the number of cells in these spheroids. 

\textbf{Size and shape of the spheroids}

\begin{figure}[h]
	\centering
	\includegraphics[width=\textwidth]{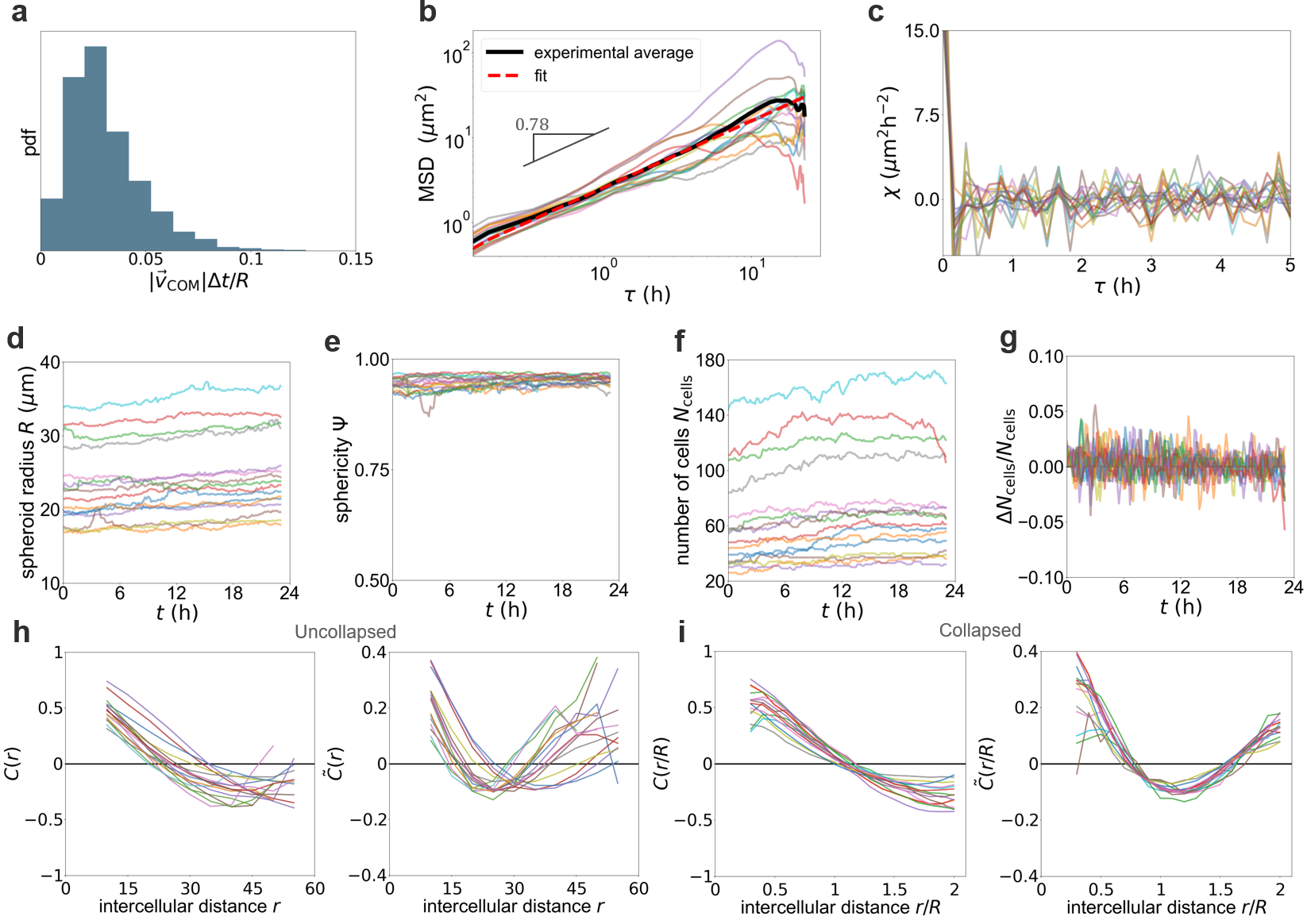}
	\caption{\textbf{Dynamics of COM motion and cell shape.} \small \textbf{a)} Distribution of the instantaneous COM displacement relative to the spheroid radius. \textbf{b)} Mean squared displacement of COM motion. Different colored lines are for individual spheroids, black line shows the average. Red dashed lines indicates the fit to the first 20 time points of the average. We fit the MSD by $C\tau^{\alpha}$ and find $C = 2.58$ and $\alpha = 0.78$. \textbf{c)} Auto correlation function of COM motion. Individual colored lines show the result of individual cell spheroids.  \textbf{d)}-\textbf{g)} Quantities characterizing the appearance of the spheroids as functions of time for different cell spheroids. d) spheroid radius $R$, e) sphericity $\Psi$, f) number of cells $N_{\text{cells}}$, g) relative number fluctuations of the number of cells $\frac{\Delta N_{\text{cells}}}{N_{\text{cells}}}$, \textbf{h),i)} Correlation function of velocity directions $C$ and velocity fluctuation directions $\tilde{C}$ as a function of the intercellular distance $r$ (h) and as a function of the rescaled intercellular distance $r/R$ (i).}
	\label{fig:spheroid_overview}
\end{figure}

We compute the radius of the spheroids by the average distance of the outermost cells of a spheroid to the COM of the spheroids. These cells are determined by creating a convex hull around the position data $\{\vec{r}_i\}$ at each point in time. Spheroid radii range from around $16 \ \mu\text{m}$ to around $37 \ \mu\text{m}$ (Fig. \ref{fig:spheroid_overview}d). To quantify the change in radius over time, we compute 
\begin{equation}
	\frac{R(t_{\text{final}}) - R(t_{0})}{R(t_{0})}
\end{equation}
where $t_{0}$ and $t_{\text{final}}$ are the beginning and final time points of the observation time period. This ratio shows that spheroid radii at the end of the observation do not vary by more than $9\%$ from their initial values at the beginning of observation. Shapes can vary over time but remain approximately spherical. We show this by considering the sphericity of the spheroids:
\begin{equation}
	\Psi = \frac{\pi^{\frac{1}{3}} (6V)^{\frac{2}{3}}}{A}
\end{equation}
The sphericity is defined by the surface area of a perfect sphere with the spheroid volume $V$ divided by the actual spheroid surface area $A$. 
Both these quantities are found from the convex hull that we construct around the spheroid. For a perfect sphere, this quantity is equal to $1$, and all other shapes have a sphericity of less than $1$. We find that the sphericity fluctuates over time but remains close to $1$ for all spheroids. This result indicates a robust spherical shape of the spheroids (Fig. \ref{fig:spheroid_overview}e), making it possible to use spherical coordinates to represent the dynamics in the surface layer of these spheroids. 

\textbf{Number of cells}

The number of cells $N_{\text{cells}}$ in the spheroids ranges from around $25$ to over $160$ for the largest spheroid (Fig. \ref{fig:spheroid_overview}f). $N_{\text{cells}}$ is not constant in time, but exhibits a small increasing upwards trend.
To further investigate this trend, we compute the rate at which the number of cells changes:
\begin{equation}
	\nu(t) = \frac{dN_{\text{cells}}}{dt}
\end{equation} 
We find that the average growth rate over all spheroids $\langle\nu(t)\rangle_{t,s} \approx 0.5 \  \text{h}^{-1}$, where $\langle\rangle_{t,s}$ indicates an average over time $t$ and different spheroid indices $s$. This positive growth rate yields a relative change of the number of cells during the measurement time $T = 23.33 \ \text{h}$, quantified by 
\begin{equation}
	\left\langle\frac{\langle\nu(t)\rangle_t T}{N_{\text{cells}}(t = 0)}\right\rangle_{s} \approx 0.2
\end{equation}
This indicates that the doubling time ($T_\text{d} \approx 5$ days) of the cell spheroids is larger than the measurement time scale ($T \approx 1$ day). Furthermore, this doubling time is much longer than the duration of a typical full rotation of a spheroid ($\approx 1$ day). To assess the instantaneous fluctuation of the number of cells defined by $\Delta N_{\text{cells}} = \nu(t)\Delta t$, we compute:
\begin{equation}
	\frac{\Delta N_{\text{cells}}}{N_{\text{cells}}} = \frac{\nu(t)\Delta t}{N_{\text{cells}}(t)}
\end{equation} 
We find that $\frac{\Delta N_{\text{cells}}}{N_{\text{cells}}}$ exhibits a narrow distribution around $0$ with maximum variations of $\pm0.04$ (Fig. \ref{fig:spheroid_overview}g). Furthermore, $\sqrt{\left\langle \left(\frac{\Delta N_{\text{cells}}}{N_{\text{cells}}}\right)^2\right\rangle} \approx 0.01$, which indicates that fluctuations of the number of cells are very small on the time scale of the measurement interval. Altogether, these results show that cell proliferation does not make an important contribution to the dynamics of the system.

%------------------------------------------------------------
\subsection{Analysis of velocity fields}

In this subsection, we elaborate on how we analyze the velocity and velocity fluctuation fields of a rotating cell spheroid whose axis of rotation is not fixed in space.

%\begin{equation}
%	\delta\vec{v}(t) = \vec{r}(t+\Delta t) - \mathcal{R}\vec{r}(t) 
%\end{equation}

%\subsubsection{Velocity- and velocity fluctuation field}

%We estimate the velocity of a cell in the COM frame of the spheroids by $\vec{v}_i = (\vec{r}_i(t+\Delta t) - \vec{r}_i(t))/\Delta t$. We define the velocity field $\vec{v}(t)$ as the collection of velocities of all cells $\{\vec{v}_i(t)\}$ at time $t$. To investigate the instantaneous migratory behavior in the rotating frame of reference, we compute the so called velocity fluctuations of a cell around the global rotation. This is done by computing $\delta\vec{v}_i(t) = (\vec{r}_i(t+\Delta t) - \mathcal{R}\vec{r}_i(t))/\Delta t$. Again we define the velocity fluctuation field $\delta\vec{v}(t)$ as the collection of velocity fluctuations of all cells $\{\delta\vec{v}_i(t)\}$ at time $t$.

\subsubsection{Correlation function scales with system size}

To identify patterns in the velocity field and in the velocity fluctuation field, we compute the spatial correlation function of velocity directions $\hat{v}(\vec{r}_i) = \vec{v}(\vec{r}_i)/|\vec{v}(\vec{r}_i)|$ and velocity fluctuation directions $\delta\hat{v}(\vec{r}_i) = \delta\vec{v}(\vec{r}_i)/|\delta\vec{v}(\vec{r}_i)|$. Note that $\vec{r}_i$ is the COM position of the i-th cell. Specifically, we compute for the velocity field \cite{Attanasi} and for the velocity fluctuation field:\\

\begin{equation}\label{eq:correlation_vel}
	C(r) = \frac{\sum_{i\neq j}^{N}\hat{v}(\vec{r}_i)\hat{v}(\vec{r}_j)\delta(\vec{r}_{ij} = r)}{\sum_{i\neq j}^{N}\delta(\vec{r}_{ij} = r)}
\end{equation}

\begin{equation}\label{eq:correlation_fluc}
	\tilde{C}(r) = \frac{\sum_{i\neq j}^{N}\delta\hat{v}(\vec{r}_i)\delta\hat{v}(\vec{r}_j)\delta(\vec{r}_{ij} = r)}{\sum_{i\neq j}^{N}\delta(\vec{r}_{ij} = r)}
\end{equation}

where we approximate the Dirac-delta function $\delta(\vec{r}_{ij} = r)$ by sharp binning in the intercellular distance $\vec{r}_{ij}$. In the main text, we claim that these correlation functions approximately collapse when we rescale the intercellular distance by the spheroid radius (Main text Fig. 1c and Main text Fig. 2a). Here we show this explicitly. The correlation functions of the velocity directions and the velocity fluctuation directions without rescaling are shown in Fig. \ref{fig:spheroid_overview}h. For these curves, we use a bin size of $dr = 5 \ \mu\text{m}$. Rescaling the intercellular distance $r$ by the spheroid radius $R$ results in a collapse of the correlation functions (Fig. \ref{fig:spheroid_overview}i). For these curves, we use a bin size of $dr = 0.1\ R$ where $R$ is the radius of the spheroids. This result suggests that the collective pattern scales with the system size on the length scales considered here (The radius of the spheroids varies between $16 \ \mu\text{m}$ to around $37 \ \mu\text{m}$).

\subsubsection{Changing perspective onto velocity fields}\label{Sec:alignment}
To analyze the dynamics of cells relative to the axis of rotation of the spheroids, we construct a transformation of the velocity and velocity fluctuation field into a frame of reference whose z-axis remains aligned with the spheroid's axis of rotation at all time points. This is done by constructing a right handed coordinate frame out of $\hat{\omega}$ as $\hat{z}'$-axis and two other orthonormal vectors where we define the $\hat{y}'$-axis to be in the plane spanned by the COM frame $\hat{z}$-axis and the axis of rotation $\hat{\omega}$. Thus, positions as well as velocity and velocity fluctuation fields are projected into this new coordinate frame. We write this transformation as a linear transformation $\mathcal{T}(t)$, equivalent to a rotation: ${\vec{r}'}(t) =\mathcal{T}(t){\vec{r}}(t)$, ${\vec{v}'}(t)=\mathcal{T}(t){\vec{v}}(t)$,  ${\delta\vec{v}'}(t)=\mathcal{T}(t){\delta\vec{v}}(t)$. Note that in the following we drop the primes again to refer to the velocity and velocity fluctuation field in the frame of reference where the axis of rotation is aligned to a new z-axis.

\subsection{Assessing the robustness of the velocity wave}\label{sec:kymo}

\begin{figure}[h]
	\centering
	\includegraphics[width=\textwidth]{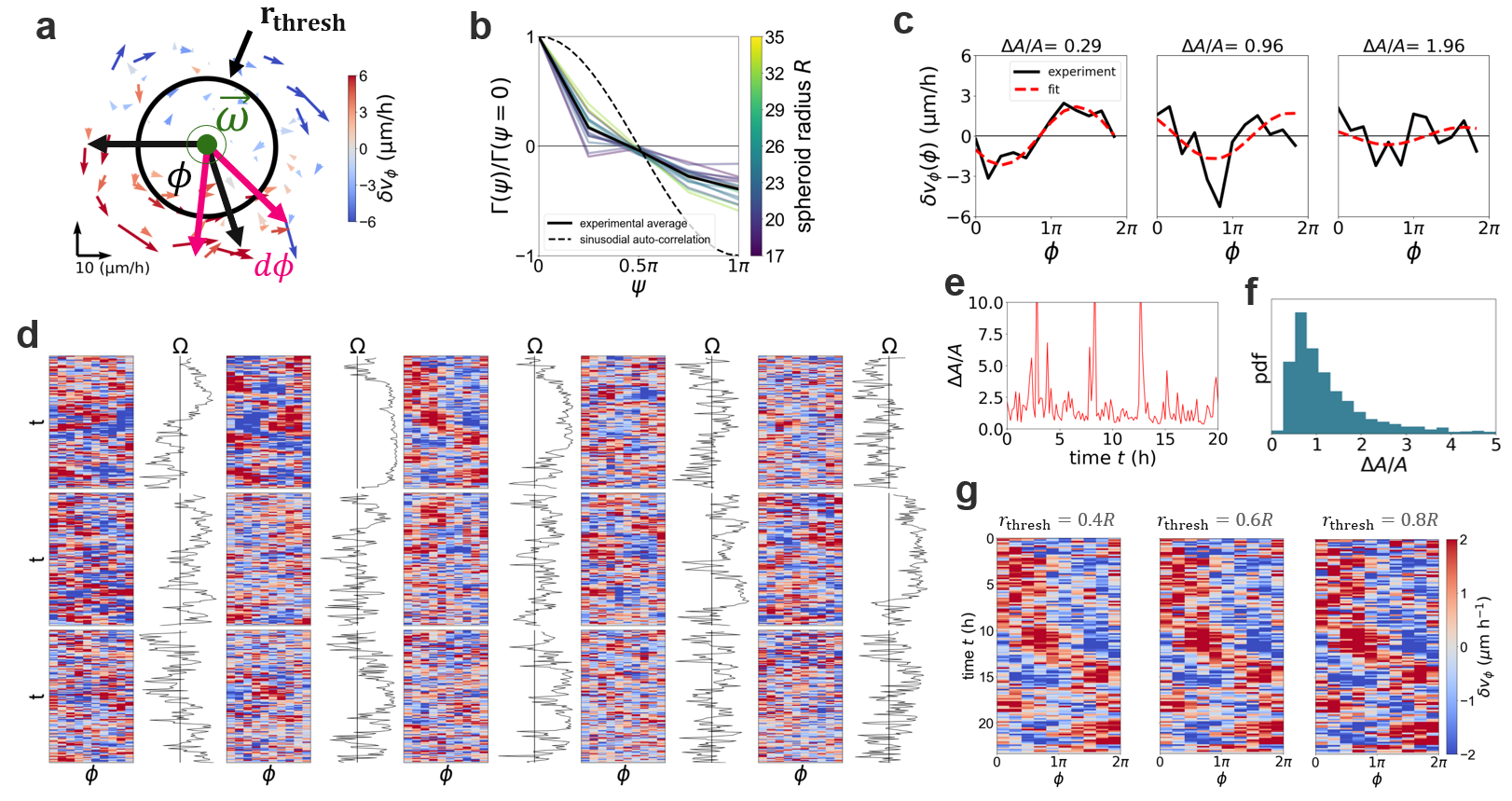}
	\caption{\textbf{Kympograph of velocity waves.} \small \textbf{a)} Schematic of the planar projections of the equatorial velocity fluctuation field. Pink vectors show margins of a bin with width $d\phi$. \textbf{b)} Auto correlation function $\Gamma(\psi)$ as defined in equation \ref{eq:autocorrelation}. Colored curves show the result for individual spheroids, black curve the experimental average and the dotted curve a sinusoidal autocorrelation. \textbf{c)} Four snapshots of the kymograph $\delta v_{\phi}(\phi,t)$ with four different noise-to-signal ratios. Black lines show the experimentally measured kymograph, red dashed line shows the fit. \textbf{d)} Kymographs of equatorial azimuthal velocity fluctuations for all cell spheroids. We use $d\phi = \pi/5$ as bin size. Always to the right we show rotational order in these spheroids. The black vertical line indicates a rotational order parameter $\Omega(t)=0.5$. Right of the black vertical lines indicates $R(t)>0.5$. \textbf{e)} Noise-to-signal ratio of one example spheroid varying over time. \textbf{f)} Distribution of noise-to-signal ratio of all cell spheroids. \textbf{g)} Kymograph of one spheroid for different values of $r_{\text{thresh}}$.}
	\label{fig:kymoraph}
\end{figure}

We observe sinusoidal velocity fluctuation profiles in the majority of the data (Fig. \ref{fig:kymoraph}a,d).
To quantify the robustness of the sinusoidal velocity fluctuation profile in the experimental data, we consider the auto-correlation function of the kymograph of equatorial velocity fluctuations:
\begin{equation*}\label{eq:autocorrelation}
	\Gamma(\psi) = \langle \delta v_{\phi}(\phi,t)\delta v_{\phi}(\phi-\psi,t)\rangle_{\phi,t}
\end{equation*}
This auto--correlation function is sinusoidal for sinusoidal velocity fluctuation profiles and can be thus used to assess the shape of $\delta v_{\phi}(\phi,t)$. We find that the experimentally measured $\Gamma(\psi)$ exhibits correlation at short distance, crosses zero at $\psi \approx 0.5\pi$ and shows anti-correlation at $\psi \approx \pi$ which corresponds to the opposite side of a spheroid (Fig. \ref{fig:kymoraph}b). This behavior is qualitatively in agreement with $\Gamma(\psi)$ for a sinusoidal profile $A(t)\sin(\phi)$. 
This shows that on average, the wave profile $\delta v_{\phi}(\phi,t)$ can be considered to be sinusoidal to a good approximation. The equatorial region of the cell spheroids thus exhibits a robust single-wavelength velocity wave profile with a wavelength approximately equal to the spheroid perimeter $\lambda_{\text{wave}} \approx 2\pi R$.\\ 
\\Throughout, we make use of the observation of a robust sinusoidal wave profile and fit the kymograph $\delta v_{\phi}(\phi,t)$ with $y_{\text{fit}}(\phi,t) = A(t)\sin(\phi - \phi_0(t))$, which yields the time dependent wave amplitude $A(t)$ and the position of the wave maximum $\phi_{\text{max}}(t) = \phi_0(t) + \pi/2$ (Fig. \ref{fig:kymoraph}c). We further consider the noise-to-signal ratio of our fit which we estimate as the normalized root-mean-square deviation  $\frac{\Delta A}{A}(t) =\sqrt{\frac{1}{N}\sum_{n=1}^{N}[\delta v_{\phi}(\phi_n,t) - y_{\text{fit}}(\phi_n,t)]^2} / A(t)$, where the sum runs over the $N$ discrete positions of the bins $\phi_n$ used for computing the kymograph. The noise-to-signal ratio of fitting the velocity fluctuation profile shows that the performance of the fitting procedure varies, indicating that sometimes the velocity fluctuation profile is less pronounced (Fig. \ref{fig:kymoraph}c,e). However, the majority of time points exhibits a noise-to-signal ratio which is close to $1$, which indicates a robust presence of the sinusoidal wave profile despite noise in the system (Fig. \ref{fig:kymoraph}f). Finally, the kymographs do not depend sensitively on decreasing $r_{\text{thresh}}$ (Fig. \ref{fig:kymoraph}g), indicating that the velocity wave is most dominant in the outer most surface layer of the spheroids.

%------------------------------------------------------------
\subsection{Wave propagation}\label{sec:wave_propagation}

In the main text, we state that the speed of wave propagation is approximately equal to the speed of the global rotation. Here, we show this explicitly. The speed of global rotation is quantified by the angular speed $\omega(t)$, which we infer from the data.  The speed of wave propagation is characterized by $\omega_{\text{wave}}(t)$, 
\begin{figure}[h]
	\centering
	\includegraphics[width=\textwidth]{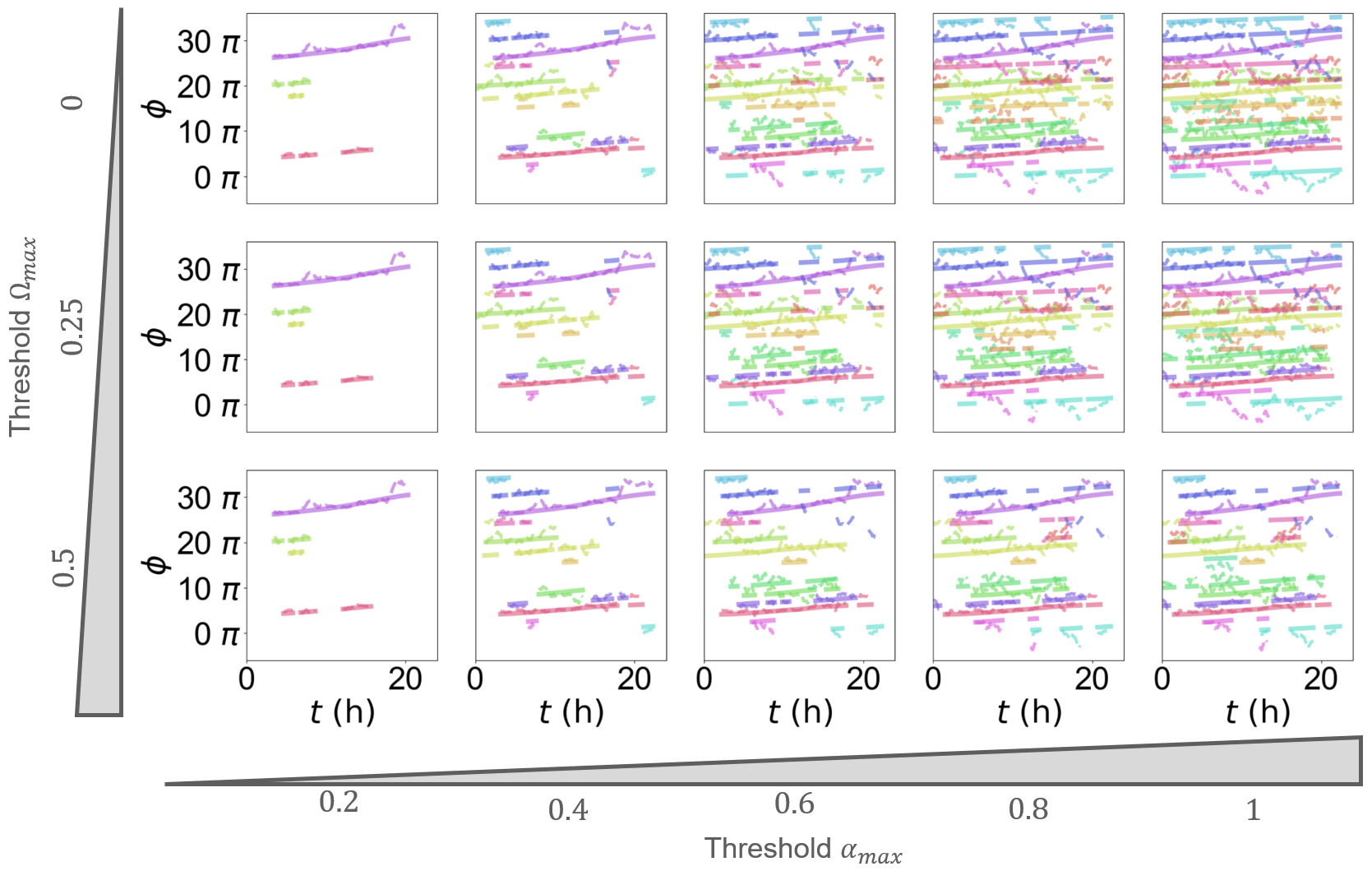}
	\caption{\textbf{Propagation of velocity wave.} \small For different values of $\alpha_{\text{max}}$ and $\Omega_{\text{max}}$ we compare the trajectory of the global rotation (solid) and the propagating velocity wave (dashed) in different cell spheroids.}
	\label{fig:wave_propagation}
\end{figure}
which we here define as the angular speed of the velocity wave maximum in the 3D COM frame. As $\omega(t)$ also quantifies the global rotation of the spheroids in the COM frame, $\omega_{\text{wave}}(t)$ is the appropriate quantity to compare to $\omega(t)$. The angular speed of the spheroids is related to the trajectory $\phi_{\text{rot}}(t) = \int_{0}^{t}\omega(t')dt'$. The wave speed $\omega_{\text{wave}}(t)$ is related to the trajectory of the wave maximum $\phi_{\text{max}}(t)$, which we obtain from the fit of a sine wave to the kymograph of equatorial velocity fluctuation profiles (section \ref{sec:kymo}). Intuitively one might expect that $\omega_{\text{wave}} = \frac{d\phi_{\text{max}}}{dt}$. However, this relation is only true if the instantaneous axis of rotation $\hat{\omega}(t) = \vec{\omega}\left(t\right)/|\vec{\omega}\left(t\right)|$ is fixed in space:  $\hat{\omega}(t) = \hat{\omega}(t+\Delta t)$. The reason for this is that the position of the velocity wave maximum $\phi_{\text{max}}(t)$ is defined with respect to the instantaneous axis of rotation (see section \ref{Sec:alignment}), not with respect to the COM of the spheroids. This means that 
$\phi_{\text{max}}(t)$ only parameterizes the 3D COM trajectory of the velocity wave (and thus the 3D COM wave speed) if the axis of rotation is fixed in space. To determine the time points where $\hat{\omega}(t) = \hat{\omega}(t+\Delta t)$ is approximately fulfilled, we quantify the movement of the axis of rotation. Specifically, we consider the angle between two subsequent instantaneous axes of rotation $\alpha(t) = \cos^{-1}(\hat{\omega}(t)\cdot\hat{\omega}(t+\Delta t))$. We define rotations as being approximately stable at time $t$ if $\alpha(t)<\alpha_{\text{max}}$ for $n$ subsequent time points where we vary $\alpha_{\text{max}}$. During these time points, we then assume $\hat{\omega}(t) \approx \hat{\omega}(t+\Delta t)$ and thus $\omega_{\text{wave}} \approx \frac{d\phi_{\text{max}}}{dt}$. Furthermore, to consider persistent global rotations, we consider time points $t$ where the rotational order parameter $\Omega(t) > \Omega_{\text{max}}$. To compare $\omega_{\text{wave}}(t)$ and $\omega(t)$, we compare the trajectories $\phi_{\text{rot}}(t)$ to $\phi_{\text{max}}(t)$ for different values of $\Omega_{\text{max}}$ and $\alpha_{\text{max}}$ (Fig. \ref{fig:wave_propagation}). In the limit of small $\alpha_{\text{max}}$, these trajectories reveal that wave propagation follows approximately the same trend as the global rotation meaning that $\omega \approx \omega_{\text{wave}}$. 

%------------------------------------------------------------
\subsection{Dynamics in the surface layer of spheroids}

In the following subsections, we show that (i) most cells reside in the surface layer, which exhibits the largest velocity fluctuations, and (ii) that cell migration is mostly tangential to the spheroid surface in this surface layer. 
\begin{figure}[h]
	\centering
	\includegraphics[width=\textwidth]{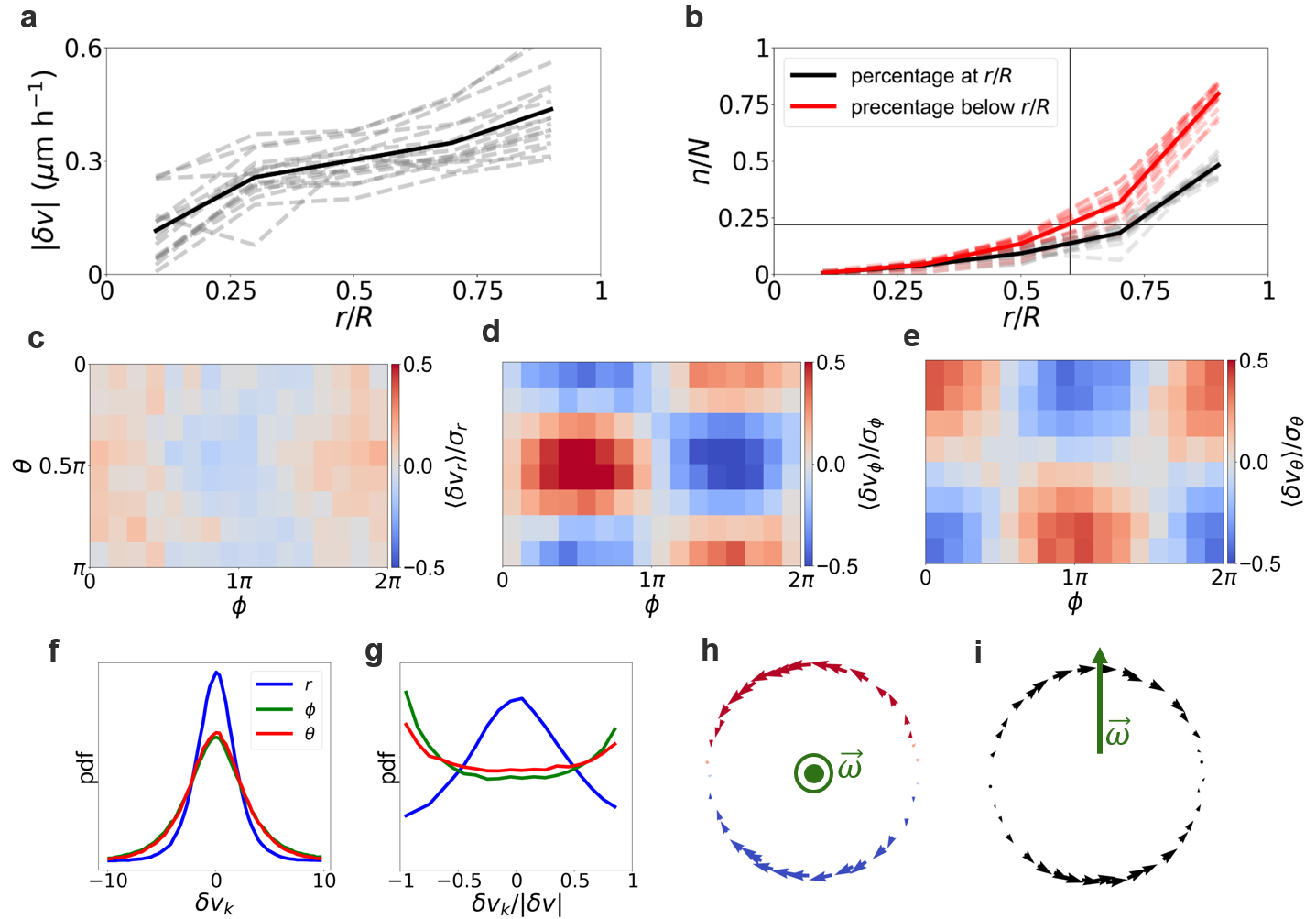}
	\caption{\textbf{Tangential surface dynamics.} \small \textbf{a)} Magnitude of velocity fluctuations as a function of the distance $r/R$ to the COM of the spheroids. Individual gray dashed curves are averages over time of individual spheroids. Black solid line indicates the average over different spheroids. \textbf{b)} The ratio of the number of cells within a shell of certain thickness at radius $r/R$ to the total number of cells in the spheroid. Again dashed lines indicate time averages of individual spheroids, while the solid line indicates the average over different spheroids. Red curves show the percentage of cells below a certain distance to the COM of the spheroids. \textbf{c)}-\textbf{e)} Noise-to-signal ratio of (c) the radial component ${\delta v}_{r}$, (d) the azimuthal component ${\delta v}_{\phi}$, and (e) the polar component ${\delta v}_{\theta}$ in the frame of reference of the propagating velocity wave. \textbf{f)} Distributions of  ${\delta v}_{\phi}$, ${\delta v}_{\theta}$, and ${\delta v}_{r}$. \textbf{g)} Distributions of ${\delta v}_{\phi}/|\delta\vec{v}|$, ${\delta v}_{\theta}/|\delta\vec{v}|$, and ${\delta v}_{r}/|\delta\vec{v}|$. \textbf{h)} Planar projection of the average velocity fluctuation field of a slab centered around the equatorial plane. \textbf{i)} Planar projection of the average velocity fluctuation field of a slice centered around the y-plane which is perpendicular to the equatorial plane.}
	\label{fig:simplify}
\end{figure}
Thus, in the main text we simplify the 3D dynamics of cell spheroids by only considering the tangential velocity fluctuations of the surface layer. This procedure then allows us to visualize the collective dynamics of the cells in the surface layer in spherical coordinates (main text Fig. 2i,k,n, Fig. 3l,m,o). In this subsection we elaborate on this approach.

\subsubsection{Importance of the surface layer for the collective dynamics}

Regarding point (i), we find that the magnitude of velocity fluctuations increases with the distance to the center of mass of the spheroids (Fig. \ref{fig:simplify}a). Furthermore, the spherical geometry implies that on average, more than $75\%$ of the cells reside beyond of $r_{\text{thresh}} = 0.6R$ (Fig. \ref{fig:simplify}b). Therefore, by focusing on the surface layer, we consider the majority of cells in the spheroids. In addition, most nearest neighbours of cells in the surface layer are also in the surface layers. In other words, cells in the surface layer are expected to interact more often among themselves than with cells in the core. Altogether, by focusing on the surface layer, we capture a majority of cells engaging in the collective dynamics of the spheroids.

\subsubsection{Radial components of the dynamics}\label{sec:tangential}

Regarding point (ii), we find that radial motion in the surface layer into or out of the spheroids is less pronounced than tangential motion in the instantaneous velocity fluctuation field. To show this, we analyze the three components of the velocity fluctuation field ${\delta v}_{\phi} = \hat{e}_{\phi}\delta\vec{v}$, ${\delta v}_{\theta} = \hat{e}_{\theta}\delta\vec{v}$, and ${\delta v}_{r} = \hat{e}_r\delta\vec{v}$ in the surface layer of cell spheroids. Not only do the two tangential components exhibit larger values than the radial component (Fig. \ref{fig:simplify} f), but they also dominate the magnitude velocity fluctuations, which we show by computing the ratio between these components and the total length of the velocity fluctuations $|\delta\vec{v}|$ (Fig. \ref{fig:simplify} g). Note that if this is the case for the velocity fluctuation field, this is also true for the velocity field, which contains the additional tangential components of the global rotation.\\
\\The small radial components in the fluctuation field do not give rise to significant average flows. To show this, we consider ${\delta v}_{\phi}$, ${\delta v}_{\theta}$, and ${\delta v}_{r}$ of the surface velocity fluctuation field in the frame of reference co-moving with the velocity wave. By coarse-graining all components through local averaging, we search for collective patterns in the directions of the surface velocity fluctuations (See section \ref{sec:averaging} for more details). We consider the signal-to-noise ratio defined by $s/\sigma = |\langle{\delta v}_{k}\rangle|/\sigma_k$, where $\sigma_k$ is the standard deviation of ${\delta v}_{k}$ and $k$ indicates the spatial components in spherical coordinates, i.e. either $r$, $\phi$, or $\theta$. As expected, ${\delta v}_{\phi}(\phi, \theta)$ and ${\delta v}_{\theta}(\phi, \theta)$ exhibit a significant pattern with four vortices (Fig. \ref{fig:simplify} d,e). However, ${\delta v}_{r}$ exhibits a signal to noise ratio which is an order of magnitude smaller than the one for the tangential components (Fig. \ref{fig:simplify}c). We thus conclude that the radial component of $\delta\vec{v}$ is not only smaller than the tangential components, but also does not exhibit significant average flows. This is in agreement with the average velocity fluctuation field, which does not show considerable radial motion (Fig. \ref{fig:simplify}h,i).

%------------------------------------------------------------
\subsection{Averaging velocity fields in the frame of reference of the velocity wave}

To characterize the tangential migratory behavior of cells in the surface layer, we find the average velocity and velocity fluctuation field in a frame of reference that is co-moving with the velocity wave. In the following section we elaborate on this averaging procedure.

\subsubsection{General procedure}\label{sec:averaging}

The average velocity and velocity fluctuation field in the frame of reference of the propagating velocity wave are found by rotating the velocity field ${\vec{v}}\left(t\right)$ and the velocity fluctuation field ${\delta\vec{v}}\left(t\right)$ to align the instantaneous axes of rotation to a new $z$-axis as described in section \ref{Sec:alignment}. 
\begin{figure}[h]
	\centering
	\includegraphics[width=\textwidth]{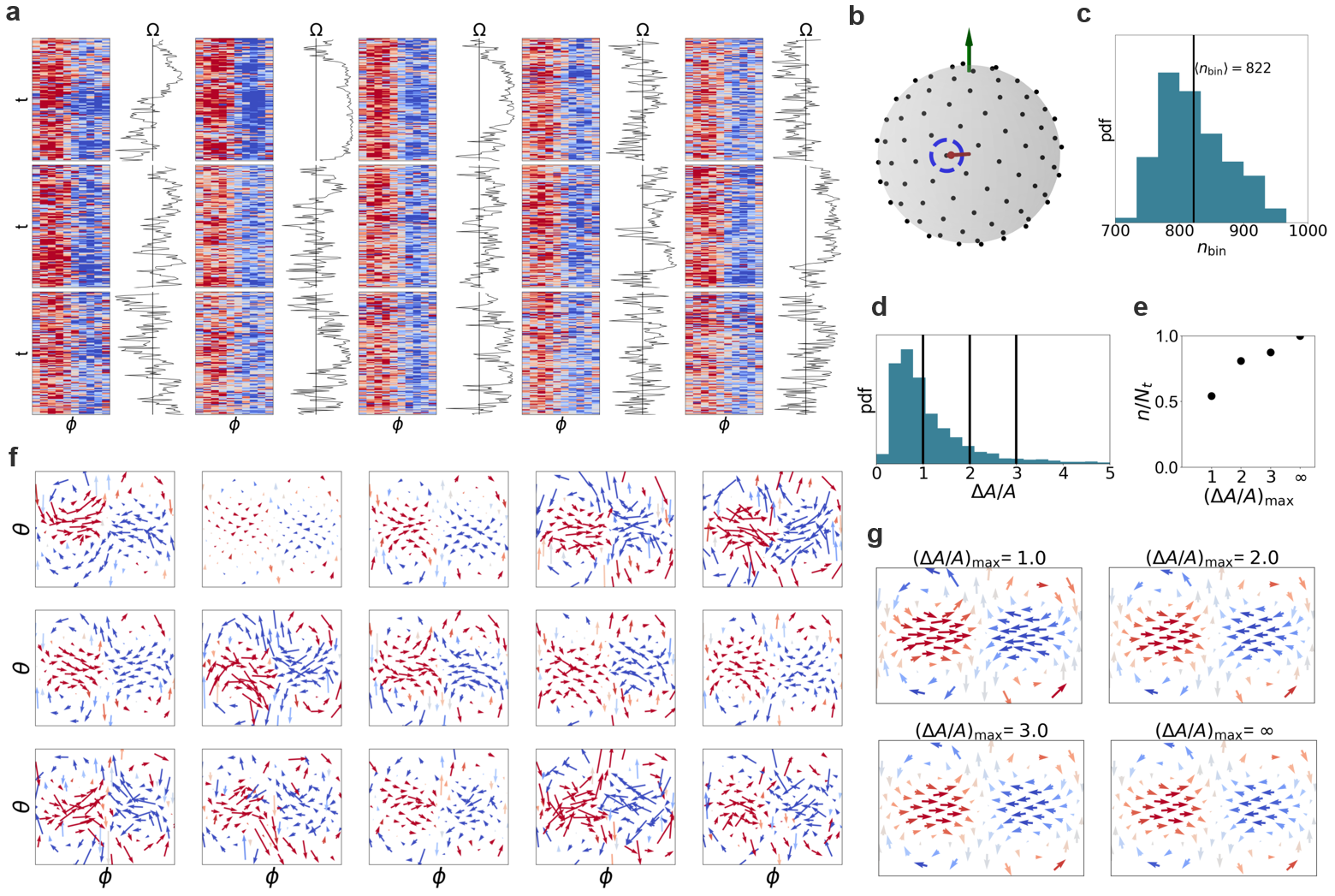}
	\caption{\textbf{Averaging procedure.} \small \textbf{a)} Kymographs of equatorial azimuthal fluctuations for many cell spheroids in the frame of reference, which is co-moving with the equatorial velocity wave. Always to the right we show rotational order in these cancer organoids. The black vertical line indicates a rotational order parameter $R(t)=0.5$. Right of the black vertical lines indicates $R(t)>0.5$. The collection of kymographs is ordered according to the radius of the organoids. \textbf{b)} Schematic of the binning of our averaging procedure. Green vector indicates the axis of rotation. Red vector indicates the position of a certain bin while the blue circle indicates the margin of this bin. \textbf{c)} Distribution of the number of cells $n_{\text{bin}}$ that end up in a bin. \textbf{d)} Distribution of noise-to-signal ratios across all cell spheroids and time points. \textbf{e)} Number of time points $n$ where $\Delta A/A < (\Delta A/A )_{\text{max}}$ divided by the total number of time points $N_t$ in the data. \textbf{f)} Overview over average tangential velocity fluctuation field averaged only over time points for individual cell spheroids. \textbf{g)} Average velocity fluctuation field in the frame of reference co-moving with the velocity wave for four different values for the threshold on the noise-to-signal ratio error $(\frac{\Delta A}{A})_{\text{max}}$.}
	\label{fig:averaging}
\end{figure}
In this frame, we compute the kymographs of the equatorial velocity wave. As these kymographs reveal the robust presence of a propagating sinusoidal wave profile, we fit the kymograph $\delta v_{\phi}(\phi,t)$ by $y_{\text{fit}}(\phi,t) = A(t)\sin(\phi - \phi_0(t))$ which yields the position of the wave maxima as detailed in section \ref{sec:kymo} and \ref{sec:wave_propagation}. To transform into the frame co-moving with the wave, we rotate the velocity field and the velocity fluctuation field around the $z$-axis so that all wave maxima are aligned at $\phi = \frac{\pi}{2}$ (Main text Fig. 2f, Fig. \ref{fig:averaging}a).  As a result of this procedure, all instantaneous velocity and velocity fluctuation fields are rotated such that both their axes of rotation and their equatorial wave maxima are aligned.\\ 
\\Then, we non-dimensionalize the velocity and velocity fluctuation fields by scaling down velocities and velocity fluctuations of each cell (indexed by $i$) by $\omega\left(t\right)r_i$, where $r_i$ is the distance of the cell to the COM. This yields ${\vec{v}}_i\left(t\right)/r_i\omega\left(t\right)$ and ${\delta\vec{v}}_i\left(t\right)/r_i\omega\left(t\right)$. Subsequently, the surface layer of the spheroids is covered in $N_{\text{bins}}=120$ uniformly distributed circular bins. The size of each bin is defined by an angle $d\Omega=0.2$ between the position vector of each bin and the position vector of a cell (Fig. \ref{fig:averaging}b). This leads to a coverage of $c = \frac{N_{\text{bins}} \pi d\Omega^2}{4\pi} = 1.2$ of the spheroids by slightly overlapping bins. All rescaled velocities and velocity fluctuations inside one bin are averaged over time and/or different experimental realizations. Averaging over all spheroids results in sufficient statistics (approx. 820 vectors per bin) (Fig. \ref{fig:averaging}c). This procedure yields an average velocity field $\langle\vec{v}/r\omega\rangle_{t,s}$ and velocity fluctuation field $\langle\delta\vec{v}/r\omega\rangle_{t,s}$. We find the tangential components of the result of this averaging procedure and represent them in spherical coordinates as in main text Fig. 2k,n and Fig. 3l,m. We show the tangential component of the average velocity fluctuation field for individual cell spheroids in Fig. \ref{fig:averaging} f.\\
\\This procedure depends on the ability to find and align sinusoidal wave profiles through fitting. During some time points mostly in smaller spheroids, we could not obtain a good fit. To assess if including these time points has an effect on the averages, we assess our fit with the noise-to-signal ratio defined as before
\begin{equation}\label{eq:noise_to_signal}
	\frac{\Delta A}{A} =\frac{\sqrt{\frac{1}{N}\sum_{n=1}^{N}[\delta v_{\phi}(\phi_n,t) - y_{\text{fit}}(\phi_n,t)]^2}}{A(t)}
\end{equation} 
To avoid including data points dominated by random fluctuations rather than a statistically significant profile, we exclude data points where noise dominates over the signal by setting a maximum threshold $(\frac{\Delta A}{A})_{\text{max}}$ in the range of observed noise-to-signal ratios (Fig. \ref{fig:averaging}d). We vary the threshold $(\frac{\Delta A}{A})_{\text{max}}$, which reveals that the large-scale pattern with four vortices in the average velocity fluctuation field is robust to changes in the threshold $(\frac{\Delta A}{A})_{\text{max}}$ (Fig. \ref{fig:averaging}g). In particular, including only data where the noise-to-signal ratio is small ($<1.0$), we observe the same characteristics of the velocity and velocity fluctuation fields as for the whole data set. Based on this insight, we thus conclude that our observations are not biased by including data points with large noise. While the large-scale features of the pattern in the velocity fluctuation field do not change with different thresholds, the amount of data which is included is of course affected by the choice of $(\frac{\Delta A}{A})_{\text{max}}$ (Fig. \ref{fig:averaging}e). In the main text, we choose $(\frac{\Delta A}{A})_{\text{max}} = 1.0$ which includes more than a half of the snapshots of the velocity and velocity fluctuation fields.

\subsubsection{Robustness of the average velocity fluctuation field}\label{sec:vorticity}

In this subsection, we demonstrate that the migratory pattern in the average velocity fluctuation showing four vortices is robustly appearing in instantaneous velocity fluctuations throughout the data. To this end, we identify vortices in the tangential velocity fluctuation field ${\delta\vec{v}}^{t}$, which we represent in spherical coordinates in main text Fig. 2i for the experimental data.\\ 

\textbf{Approximating the vorticity field}

To locate the vortices in the velocity fluctuation field, we compute a proxy for the vorticity field of the tangential velocity fluctuation field $\vec{\nabla}\times\delta\vec{v}^{t}$ by the use of Stoke's theorem. Specifically, we relate the flow tangential to a circle $\mathcal{C}({\vec{r}}, R_{\mathcal{C}})$ centered around a certain position ${\vec{r}}$ with a radius $R_{\mathcal{C}}$ to the vorticity of the tangential velocity fluctuation field inside the disk $\mathcal{D}({\vec{r}}, R_{\mathcal{C}})$ which the circle encloses:
\begin{equation}
	\int_{\mathcal{D}({\vec{r}}, R_{\mathcal{C}})} (\vec{\nabla}\times\delta\vec{v}^{t}) dV = \oint_{\mathcal{C}({\vec{r}}, R_{\mathcal{C}})}\delta\vec{v}^{t}\cdot d\hat{l} = \Upsilon\left({\vec{r}}\right)
\end{equation}
Here, the vector $d\hat{l}$ is a unit vector that is tangential to the circle as well as to the spheroid surface. Integrating along the circle yields the line integral of $\delta\vec{v}^{t}$ along the circle. We aim to find the value of this integral, which we call $\Upsilon\left(\vec{r}\right)$. This quantity acts as a proxy to the vorticity and can be interpreted as a local “angular momentum” of the velocity fluctuation field around a certain position ${\vec{r}}$ parameterized by $\left(\phi,\theta\right)$:
\begin{equation}
	{\vec{r}}={R\left(\sin{\left(\theta\right)}\cos{\left(\phi\right)},\sin{\left(\theta\right)}\sin{\left(\phi\right)},\cos{\left(\theta\right)}\right)}^T
\end{equation} 

\begin{figure}[h]
	\centering
	\includegraphics[width=\textwidth]{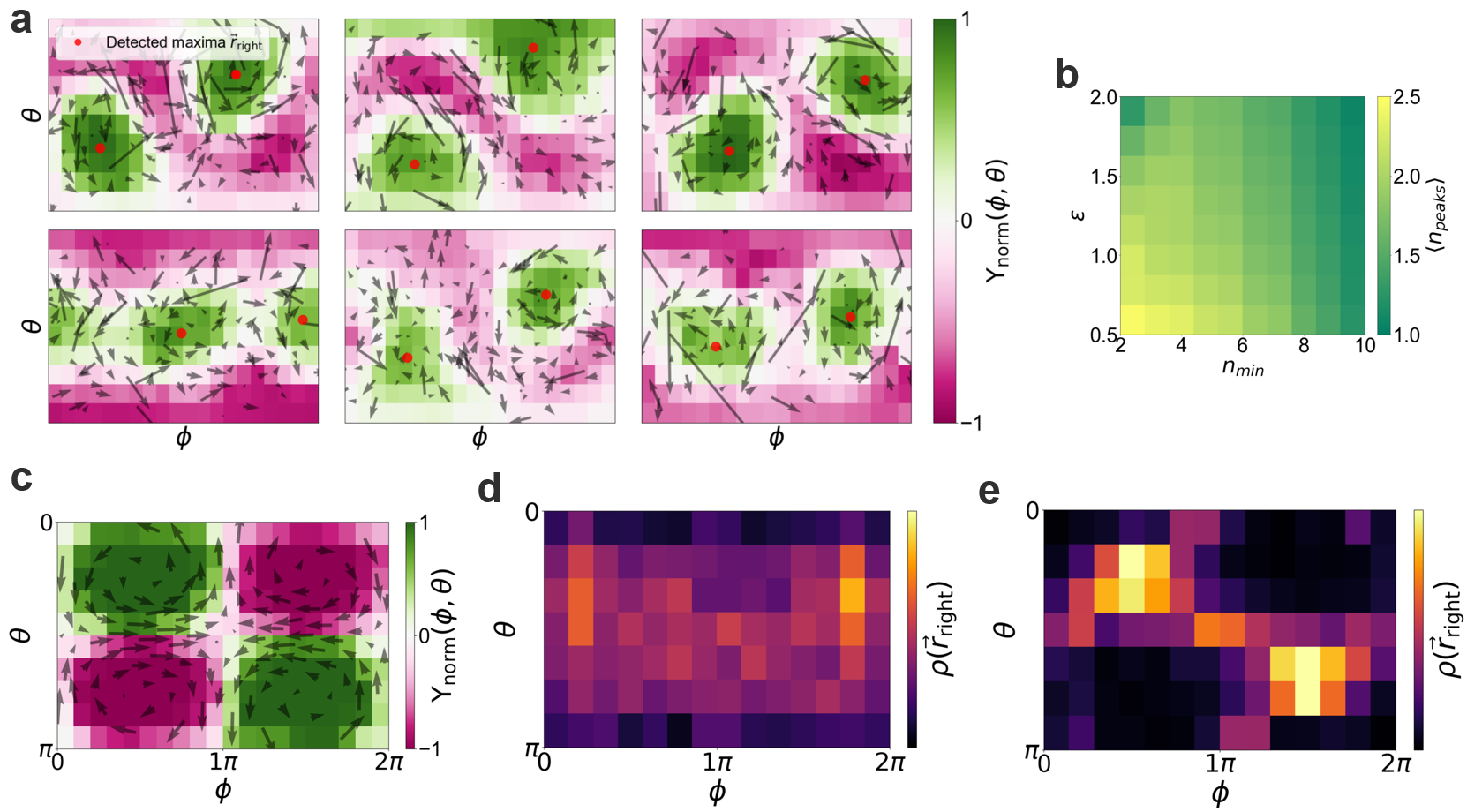}
	\caption{\textbf{Robustness of the supracellular pattern.} \small \textbf{a)} Snapshots of the tangential velocity fluctuation field and the normalized vorticity measure $\Upsilon\left(\phi,\theta\right)$. \textbf{b)} Average number of right handed vortices in the velocity fluctuation field. We vary the two hyper parameters within a reasonable regime dictated by the typical distance of the data points. \textbf{c)} $\Upsilon\left(\phi,\theta\right)$ of the average velocity fluctuation field. \textbf{d)},\textbf{e)} Distribution of the positions of the right-handed swirls  $\vec{r}_{\text{right}}$ in the COM frame (d) and in the frame co-moving with the velocity wave (e).}
	\label{fig:robustness_swirls}
\end{figure}

To compute $\Upsilon\left(\phi,\theta\right)$ from discrete data, we express the circle radius as $R_{\mathcal{C}} = R\Omega$ where $R$ is the spheroid radius and $\Omega$ is an angle between ${\vec{r}}$ and the positions of the cells $\{\vec{r}_i\}$. We approximate the circle by constraining the angle between ${\vec{r}}$ and the positions of the cells $\{\vec{r}_i\}$ in the bin $[\Omega-d\Omega,\Omega+d\Omega]$, where we choose $\Omega = 0.8$ and $d\Omega = 0.4$. Furthermore, we focus on the surface layer so we constrain $|\vec{r}_i| > 0.6R$. Then all projections $\delta\vec{v}^{t}_{i}\hat{l}_i$ along the circle are summed up to approximate the integral: 
\begin{equation}
	\Upsilon\left(\phi,\theta\right) \approx \sum_{i=1}^{M} \delta\vec{v}^{t}_{i}\cdot\hat{l}_i
\end{equation}
Here $\hat{l}_i$ is computed at the position $\vec{r}_i$ of the $i$-th cell in the bin with $M$ cells total being close to the circle. In the following, we consider the 'normalized angular momentum' defined by  
\begin{equation}
	\Upsilon_{\text{norm}}\left(\phi,\theta\right) = \frac{\sum_{i=1}^{M} \delta\vec{v}^{t}_{i}\cdot\hat{l}_i}{\sum_{i=1}^{M} |\delta\vec{v}^{t}_{i}\cdot\hat{l}_i|}
\end{equation}
This normalized quantity takes values of $\Upsilon_{\text{norm}}\left(\phi,\theta\right)\in[-1,+1]$, where $+1$ indicates perfect counter clock-wise rotation, and $-1$ indicates perfect clock-wise rotation as seen from along ${\vec{r}}$ towards the center of the spheroids.

\textbf{Patterns in the vorticity field}

Snapshots of $\Upsilon_{\text{norm}}\left(\phi,\theta\right)$ reveal that the pattern with four vortices found in the average (Fig. \ref{fig:robustness_swirls}d) is prominent throughout the data (Fig. \ref{fig:robustness_swirls}a). To quantify this robustness, we locate the maxima in $\Upsilon_{\text{norm}}\left(\phi,\theta\right)$, which yields the positions of the right handed vortices $\vec{r}_{\text{right}}$. We do so by finding the positions $\left(\phi,\theta\right)$ where $\Upsilon_{\text{norm}}\left(\phi,\theta\right) > 0.5$. These positions are then clustered by a clustering algorithm called \textit{DBSCAN}. This algorithm identifies clusters dependent on the local maxima in density of data points using two hyper parameters $\varepsilon$ and $n_{\text{min}}$ \cite{10.5555/3001460.3001507}. We average the positions within one cluster to find the average position of the local maxima. Within a range of hyper parameters, we find that the velocity fluctuation field exhibits around two right handed vortices (Fig. \ref{fig:robustness_swirls}b), which is in agreement with the average velocity fluctuation field (Fig. \ref{fig:robustness_swirls}d). Plotting the positions of these vortices in the COM frame, leads to a flat probability distribution of their positions (Fig. \ref{fig:robustness_swirls}d), indicating that vortices propagate. However, the transformation into the frame of reference co-moving with the velocity wave, also aligns the positions of the vortices (Fig. \ref{fig:robustness_swirls}e). This shows that transforming into a frame of reference co-moving with the velocity wave can be regarded as equivalent to transforming into a frame of reference co-moving with the four vortices. Thus, there is a robust coupling between the velocity wave and the accompanying vortices, showing that the four vortices in the average velocity fluctuation field accurately represent the migratory dynamics at individual time points. 

\subsection{Density modulations}\label{sec:density}

\begin{figure}[h]
	\centering
	\includegraphics[width=\textwidth]{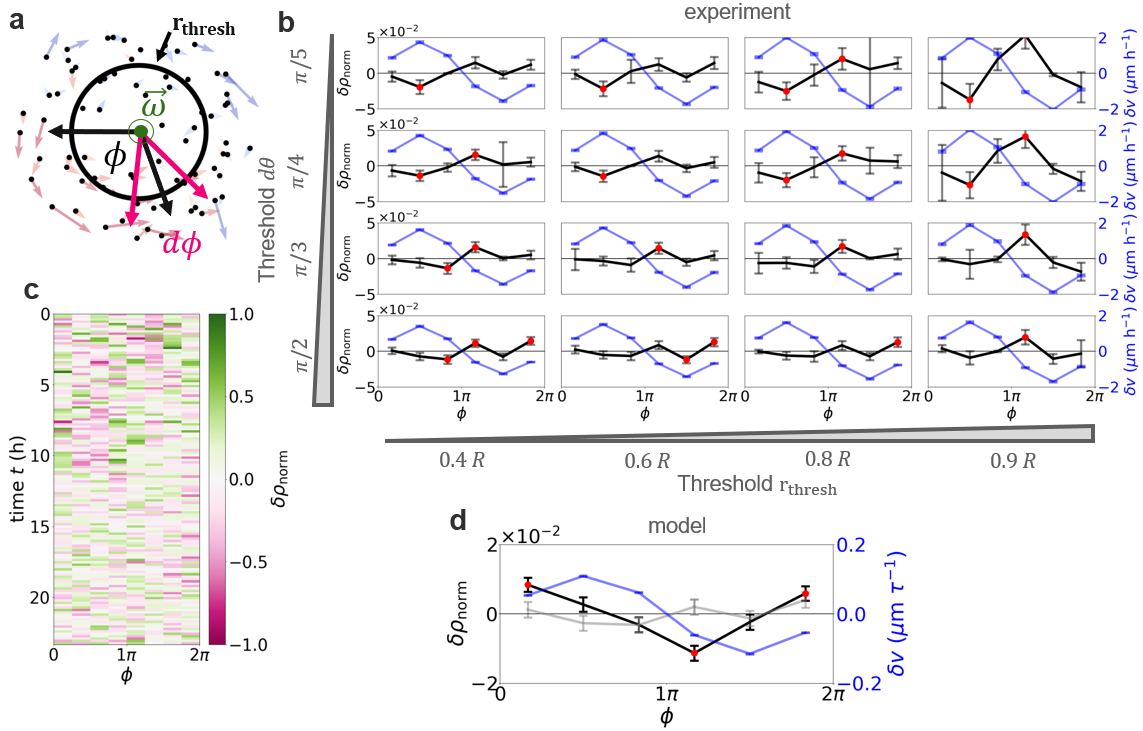}
	\caption{\textbf{Density modulations.} \small \textbf{a)} Schematic of the procedure with which we measure density modulations in the cell spheroids. Black dots indicate the positions of the cells. \textbf{b)} Overview over the density fluctuation $\langle\delta\rho_{\text{norm}}(\phi,t)\rangle_{t,s}$ averaged over time and different spheroids for different values of the parameters $d\theta$ and $r_{\text{thresh}}$ (black curves). Red data points are significantly different from zero (p-value $< 0.05$) according to a two-sided t-test. Blue curves show the equatorial velocity wave profile. Here, we average over time and different cell spheroids (without making a significance test). Error bars for both the density fluctuation profile and the velocity fluctuation profile represent the standard error of the mean (s.e.m) computed by bootstrapping.  \textbf{c)} Kymograph of the normalized surface density in the frame co-moving with the velocity wave with $d\phi = \pi/4$. \textbf{d)} Time-averaged normalized surface density $\delta\rho_{\text{norm}}(\phi)$ for $\epsilon/\gamma = 0.8 \tau^{-1}$ (black) and $\epsilon/\gamma = 5 \tau^{-1}$ (gray), and velocity wave $\delta v(\phi)$ (blue) in the frame co-moving with the velocity wave in our model. Again error bars for both the density fluctuation profile and the velocity fluctuation profile represent the standard error of the mean (s.e.m) computed by bootstrapping.}
	\label{fig:density}
\end{figure}

In this subsection, we elaborate on how we investigate the cell density along the velocity wave in the equator of the cell spheroids. To this end, we build a kymograph for the cell density following the procedure outlined in section \ref{sec:kymo}, but apply it to the cell density: We count the number of cells inside the bins constrained to the equatorial region by $\theta\in[\frac{\pi}{2} - d\theta, \frac{\pi}{2} + d\theta]$, to the surface layer by $r_{\text{thresh}} = 0.6R$, and within the bin of width $d\phi$ in which we found the average azimuthal component of the velocity fluctuation (Fig. \ref{fig:density}a). 
To approximate the surface density of cells we divide the number of cells in these bins by the surface area of the bins $V = d\phi R^2 \left(-\cos(\pi/2 + d\theta) - \cos(\pi/2 - d\theta)\right)$. Complementary to the kymograph containing the velocity wave $\delta v(\phi,t)$, we thus obtain for all spheroids the kymograph of cell surface density $\rho_{\text{s}}(\phi,t)$. We consider the density fluctuations:
\begin{equation}
	\delta\rho_{\text{s}}(\phi,t) = \rho_{\text{s}}(\phi,t) - \langle\rho_{\text{s}}(\phi,t)\rangle_{\phi}
\end{equation} 
where we use the average surface density in the equator $\langle\rho_{\text{s}}(\phi,t)\rangle_{\phi}$. Furthermore, we consider the normalized density fluctuations:
\begin{equation}
	\delta\rho_{\text{norm}}(\phi,t) = \frac{\rho_{\text{s}}(\phi,t) - \langle\rho_{\text{s}}(\phi,t)\rangle_{\phi}}{\langle\rho_{\text{s}}(\phi,t)\rangle_{\phi}}
\end{equation} 

We ask if the equatorial density profile in the frame co-moving with the velocity wave exhibits significant modulations. To this end, we consider the kymograph of the normalized density fluctuations $\delta\rho_{\text{norm}}(\phi,t)$ in the frame of reference of the moving velocity wave. Here, we do not observe a large scale density profile synchronized to the velocity wave (compare Fig. \ref{fig:density}c to main text Fig. 2f). We nevertheless test if the kymograph we found is significantly different from a uniform profile. To this end, we consider the time and spheroid average $\langle\delta\rho_{\text{norm}}(\phi,t)\rangle_{t,s}$ for different values of the threshold $d\theta$ which separates the equatorial region from the polar region as well as the threshold $r_{\text{thresh}}$ which separates the core of the spheroids from the surface layer (Fig. \ref{fig:density}b). To test for significance, we perform a two-sided t-test to determine whether the density fluctuations are significantly different from $0$. First, we find that the density fluctuations are small with respect to the mean: $\langle\delta\rho_{\text{norm}}(\phi,t)\rangle_{t,s} \sim 0.01$. Secondly, while varying the parameters $d\theta$ and $r_{\text{thresh}}$, we do not find a robust profile of $\langle\delta\rho_{\text{norm}}(\phi,t)\rangle_{t,s}$ with regions that are significantly different from $0$ in general in the frame of reference of the velocity wave. In contrast, the velocity wave profile we obtain by the same procedure is robustly significant (blue curves in Fig. \ref{fig:density}b). We thus conclude that in the frame of reference of the propagating velocity wave, density modulations are not only small but also statistically not significant. These results suggest that the velocity wave in rotating cell spheroids is not coupled to a density wave.

\subsection{Flux analysis}

\begin{figure}[h]
	\centering
	\includegraphics[width=\textwidth]{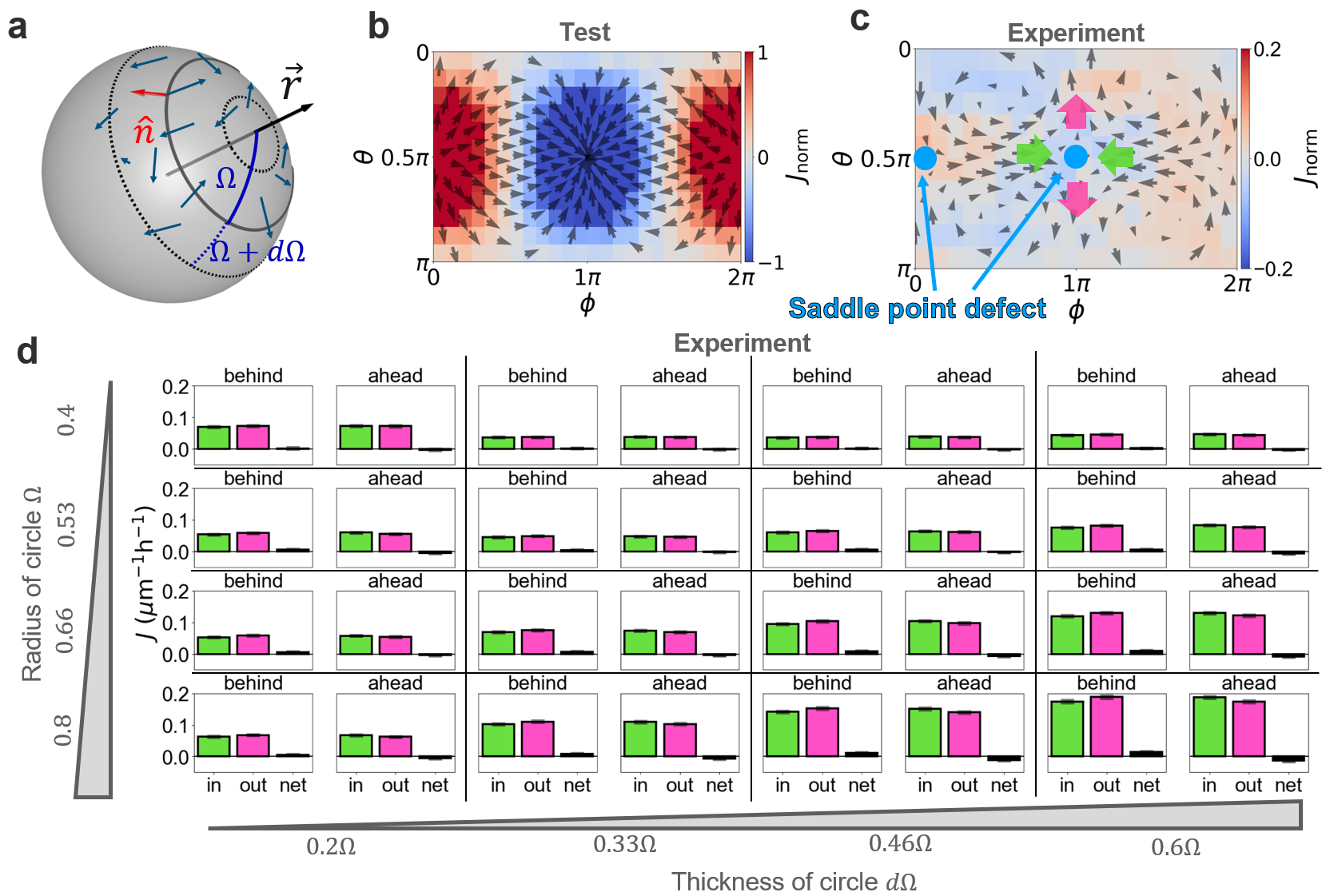}
	\caption{\textbf{Analysis of fluxes.} \small \textbf{a)} Schematic of measuring cell fluxes along the spheroid surface. Blue vectors indicate a tangential vector field. The angle $\Omega$ defines the radius of a circle, $d\Omega$ defines the thickness of the circle, which is a parameter that is needed to compute fluxes through a surface from finite data. \textbf{b)} The normalized cell flux $J_{\text{norm}}\left(\phi,\theta\right)$ for a test vector field with one source and one sink. \textbf{c)} The normalized cell flux $J_{\text{norm}}\left(\phi,\theta\right)$ for the experimental data in the frame moving with the velocity wave. We average over time and different spheroids as we do to find the average velocity fluctuation field, which is shown in front. \textbf{d)} Quantitative analysis of the time- and spheroid averaged net cell flux $J_{\text{net}}$ and the time- and spheroid averaged in- and out flux $J_{\text{in/out}}$ at the positions of the two saddle-point defects ahead and behind the velocity wave (See panel c). For the influx and outflux, we show the absolute value. We vary the parameters $\Omega$ and $d\Omega$ to reveal how sensitive our analysis is with respect to the choice of parameters. Error bars are barely visible but represent the standard error of the mean (s.e.m) computed by bootstrapping.}
	\label{fig:Flux_analysis}
\end{figure}

We observe that cells at the saddle-point defects behind and ahead of the velocity wave maximum divert towards the poles. To investigate whether this motion is of incompressible nature, we analyze the cell flux around these defects. In particular, as we show that tangential motion is most dominant in the surface layer of the spheroids (Fig. \ref{fig:simplify}c-i), we consider the instantaneous tangential cell flux along the spheroid surface $\vec{j} = \rho_s\delta\vec{v}^{t}$, where $\rho_s$ is the surface density of cells in the surface layer of the spheroids and $\delta\vec{v}^{t}$ is the tangential component of the velocity fluctuation field, which exhibits two saddle-point defects and four vortex defects. Importantly, note that in the following we analyze the fluxes in a frame of reference which is co-moving with the velocity wave and thus also with the vortex defects in the velocity fluctuation field (Fig. \ref{fig:robustness_swirls}e). Therefore, the saddle-point defects are on average located at the same positions in this frame. We found no significant density modulation along the equator of the rotating spheroids. Therefore, any net flux through a circle at the surface of the spheroid is related to divergences stemming from cell flux in the radial direction. To assess this possibility, we aim to find the net flux through a circle $\mathcal{C}({\vec{r}}, R_{\mathcal{C}})$ centered around a certain position ${\vec{r}}$ with a radius $R_{\mathcal{C}}$:
\begin{equation}
	J_{\text{net}}\left({\vec{r}}\right) = \oint_{\mathcal{C}({\vec{r}}, R_{\mathcal{C}})}\rho_s\delta\vec{v}^{t}\cdot d\hat{n}
\end{equation}
Here, the vector $d\hat{n}$ is a unit vector that is normal to the circle, pointing away from the center of the circle, and is also tangential to the spheroid surface (Fig. \ref{fig:Flux_analysis}a). To compute $J_{\text{net}}\left({\vec{r}}\right)$ from discrete data, we consider a circular cap on the surface of the sphere centered around a certain position ${\vec{r}}$, which is parameterized by $\left(\phi,\theta\right)$:
\begin{equation}
	{\vec{r}}={R\left(\sin{\left(\theta\right)}\cos{\left(\phi\right)},\sin{\left(\theta\right)}\sin{\left(\phi\right)},\cos{\left(\theta\right)}\right)}^T
\end{equation} 
We express the circle radius as $R_{\mathcal{C}} = R\Omega$ where $R$ is the spheroid radius and $\Omega$ is an angle between ${\vec{r}}$ and the positions of the cells $\{\vec{r}_i\}$ (Fig. \ref{fig:Flux_analysis}a). We approximate the circle by constraining the angle between ${\vec{r}}$ and the positions of the cells $\{\vec{r}_i\}$ in the bin $[\Omega-d\Omega,\Omega+d\Omega]$, where we vary $\Omega \in [0.4, 0.8]$ and $d\Omega \in [0.2\Omega, 0.6\Omega]$. 

Furthermore, we focus on the surface layer so we constrain $|\vec{r}_i| > 0.6R$. Then all projections $\rho_{s}^{i}\delta\vec{v}^{t}_{i}\cdot\hat{n}_i$ inside the circular bin are summed up to approximate the integral: 
\begin{equation}
	J_{\text{net}}\left(\phi,\theta\right) \approx \sum_{i=1}^{M} \rho_{s}^{i}\delta\vec{v}^{t}_{i}\cdot\hat{n}_i
\end{equation}
Here $\hat{n}_i$ is computed at the position $\vec{r}_i$ of the $i$-th cell in the bin with $M$ cells. Furthermore, $\rho_{s}^{i}$ is the local surface density of the spheroid at the position of the $i$-th cell. We approximate this density by considering the close vicinity of the $i$-th cell defined by an angle $\Omega_{\rho_s} = 0.8$ between this cell and surrounding cells. We count the number of cells inside this vicinity and divide it by its area, which be approximate as that of a flat disk. This local approximation yields values close to the ones we measure for the global surface density in section \ref{sec:density}. We then also consider the influx and outflux: 
\begin{equation}
	J_{\text{in/out}}\left(\phi,\theta\right) = \sum_{i=1, \delta\vec{v}^{t}_{i}\cdot\hat{n}_i\gtrless0}^{M} \rho_{s}^{i}\delta\vec{v}^{t}_{i}\cdot\hat{n}_i
\end{equation}

and the normalized net flux through the circle boundary: 
\begin{equation}
	J_{\text{norm}}\left(\phi,\theta\right) = \frac{\sum_{i=1}^{M} \rho_{s}^{i}\delta\vec{v}^{t}_{i}\cdot\hat{n}_i}{\sum_{i=1}^{M} |\rho_{s}^{i}\delta\vec{v}^{t}_{i}\cdot\hat{n}_i|}
\end{equation}
This normalized quantity takes values of $J_{\text{norm}}\left(\phi,\theta\right)\in[-1,+1]$, where $+1$ indicates pure influx while $-1$ indicates pure outflux. We test our analysis by computing the relative flux for an analytically given tangential vector field that features two fully divergent aster defects (Fig. \ref{fig:Flux_analysis}b).\\
\\For the experimental data, we find that the average over time and different spheroids of $J_{\text{norm}}\left(\phi,\theta\right)$ is in general close to $0$ in the frame of reference of the velocity wave. This indicates that on average the majority of cells that tangentially flow through a certain circular region of the surface layer, do this in a way that cell fluxes are approximately balanced (Fig. \ref{fig:Flux_analysis}c). We further analyze the tangential cell flux around the saddle-point defects. We find that on average also in these regions the cell influx and the cell outflux are approximately balanced (Fig. \ref{fig:Flux_analysis}d). Therefore, from the absence of significant density modulations and the balance of fluxes, we conclude that the velocity fluctuation field describes incompressible tangential cell motion along the spheroid surface. 

\section{Stochastic rigid-body rotation}

The correlation function of experimental velocity directions shown in main text Fig. 1c reveals alignment of nearest neighbors and anti-alignment of cells that are on opposite sides of the spheroid. This correlation structure indicates a robust mode of global rotation. Therefore, we consider a stochastic rigid-body rotation as minimal model for the rotational dynamics of the cell spheroids. This model features deterministic rotation with a constant angular velocity everywhere in space and uncorrelated Gaussian white noise. We implement this model via the following Euler-forward scheme:
\begin{equation}
	\vec{r}(t+\Delta t) = \mathcal{R}(t)\vec{r}(t) + \sqrt{\Delta t}\vec{\eta}(t)
\end{equation}
\begin{equation}
	\langle\eta_i(t)\eta_j(t')\rangle = \sigma^2 \delta_{ij}\delta(t-t')
\end{equation}
Here, $\mathcal{R}(t)$ is a rotation matrix and $\vec{\eta}(t)$ is uncorrelated Gaussian white noise with amplitude $\sigma$. To facilitate the comparison to the experiment, we use the inferred rotation matrix $\mathcal{R}(t)$ from the experimental cell spheroids in our model. Furthermore, assuming that the dynamics beyond the global rotation of the experimental cell spheroids can be described by uncorrelated Gaussian white noise, we determine the noise amplitude at time $t$ from the experimental velocity fluctuations $\delta\vec{v}$ according to 
\begin{equation}\label{eq:noise_amplitude}
	3\sigma^2 = \langle\vec{\eta}(t)^2\rangle = \frac{1}{\Delta t}\langle(\vec{r}(t+\Delta t) - \mathcal{R}(t)\vec{r}(t))^2\rangle = \Delta t\langle\delta\vec{v}^2\rangle
\end{equation}
Importantly, we observe that $|\delta\vec{v}|$ increases with the distance of the cells to the spheroid's COM (Fig. \ref{fig:simplify}a). 
\begin{figure}[h]
	\centering
	\includegraphics[width=\textwidth]{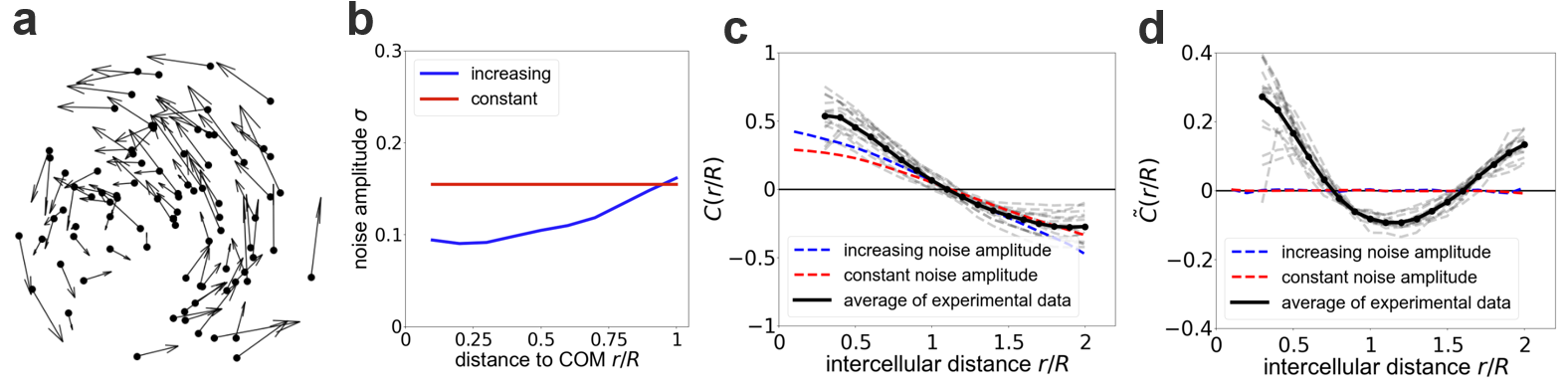}
	\caption{\textbf{Stochastic rigid body rotation.} \small \textbf{a)} Snapshot of a velocity field predicted by the rigid body rotation model. \textbf{b)} Noise amplitude inferred from the experimental data. \textbf{c)},\textbf{d)} Correlation of velocity (c) and velocity fluctuation (d) directions compared between the experiment (black), rigid body rotation with constant noise amplitude (red) and rigid body rotation with increasing noise amplitude (blue).}
	\label{fig:rigid_body_rotation}
\end{figure}
To capture this radius dependency in the minimal stochastic rigid-body model, we use a space-dependent noise amplitude $\sigma(r)$, where $r$ is the distance of a cell to the COM of the spheroid. We compute this noise amplitude from the experimental data using the following conditional average \cite{Brueckner}:
\begin{equation}\label{eq:noise_amplitude_radius}
	\sigma(r) = \sqrt{\frac{\Delta t}{3}\langle\delta\vec{v}^2||\vec{r}| \approx r\rangle}
\end{equation}
We simulate $16$ spherical clouds of point particles with the size of the $16$ cell spheroids and approximate number of cells in the spheroids (Fig. \ref{fig:rigid_body_rotation}a). We choose the same time scale as in the experiment: $\Delta t = 10 \text{min}$ and $140$ frames. From the simulations, we then compute the correlation function of velocity directions and velocity fluctuation directions using equations \ref{eq:correlation_vel} and \ref{eq:correlation_fluc}. The correlation of velocity directions shows alignment at small intercellular distances and anti-alignment at larger intercellular distances for both a constant noise amplitude and a space-dependent noise amplitude. This result shows that the overall spatial correlations of the velocity field of the global rotational dynamics of cell spheroids can be approximately captured by a stochastic rigid-body rotation. Note that using a constant noise amplitude for the stochastic rigid-body rotation underestimates the correlation at small distances while the space-dependent noise amplitude captures the experimental velocity direction correlation function more quantitatively (Fig. \ref{fig:rigid_body_rotation}b,c).\\
\\In contrast, the correlation of velocity fluctuation directions in our stochastic rigid-body model vanishes, which shows no coordination of the velocity fluctuations of this simplified model (Fig. \ref{fig:rigid_body_rotation}d). This is by construction as we impose only uncorrelated noise in addition to the deterministic rotation. This result is in clear disagreement with the experimental data which shows pronounced non-monotonic correlation of velocity fluctuation directions (Main text fig. 2a). This shows that the simplified model of stochastic rigid body rotation is not sufficient to capture the velocity waves that we discover in rotating cell spheroids. 

\newpage

%------------------------------------------------------------
\section{Model Implementation}\label{sec:model}

We employ a minimal active particle model commonly used to model collective cell migration \cite{PhysRevE.74.061908, Copenhageneaar8483, Sepulveda, Alessandro}. Specifically, we model cells as overdamped particles moving on a substrate with effective friction coefficient $\gamma$ while being self-propelled into the direction of their internal polarization $\hat{p}$ with speed $v_0$. The polarization itself is subject to rotational noise with amplitude $\sigma$ and exhibits alignment interactions with strength $\beta$. The excluded volume interactions are modeled by the repulsive part of a linear force $\vec{F}_{\text{rep}}^{i}(r_{ij})$ with amplitude $\varepsilon$ parameterizing the stiffness of the soft particles. By this repulsion interaction, the particle has an effective radius $\lambda$ above which we cut off the attractive force. Following \cite{Sknepnek}, we write the equations of motion for these self-propelled particles constrained to a spherical surface as:
\begin{equation}\label{eq:v_projected}
	\vec{v}_i = \mathbf{P}_{T}(v_0\hat{p}_i(t) + \frac{1}{\gamma}\vec{F}_{\text{rep}}^{i}(r_{ij}))
\end{equation}
\begin{equation}\label{eq:repulsion_projected}
	\vec{F}^{i}_{rep}(r_{ij}) = -\varepsilon\sum_{r_{ij} < r_{\text{inter}}}(2\lambda - r_{ij})\hat{r}_{ij} 
\end{equation}
\begin{equation}\label{eq:polarity_projected_rotation}
	\frac{d\hat{p}_i}{dt} = \dot{\phi}_i(\hat{r}_i(t) \times \hat{p}_i(t)) 
\end{equation}
\begin{equation}\label{eq:polarity_projected}
	\dot{\phi}_i = -\beta\sum_{r_{ij} < r_{\text{inter}}}\sin(\phi_i(t) - \phi_j(t)) + \eta_i(t) =  -\beta \sum_{r_{ij} < r_{\text{inter}}}(\hat{p}_i(t) \times \hat{p}_j(t))\cdot\hat{r}_i(t)  + \eta_i(t) 
\end{equation}
\begin{equation}\label{eq:noise}
	\langle \eta_i(t)\eta_j(t')\rangle = \sigma^2 \delta(t-t')\delta_{ij} 
\end{equation} 
$\mathbf{P}_{T}$ in equation \ref{eq:v_projected} is a projection operator that projects the forces onto the tangential plane of particle $i$ at position $\vec{r}_i$ \cite{Sknepnek}. Equation \ref{eq:repulsion_projected} describes the repulsion interaction and contains the inter-particle distance $r_{ij}$ and a vector $\hat{r}_{ij}$ that points from particle $i$ to $j$. Here, the sum runs over all cells $j$ within a radius of interaction of cell $i$ $r_{ij} < r_{\text{inter}}$. Equation \ref{eq:polarity_projected_rotation} describes the dynamics of a unit vector that can only rotate in the tangential plane around the normal radial vector $\hat{r}_i$ at the position of the $i$-th particle $\vec{r}_i$. The angle of rotation is determined by equations \ref{eq:polarity_projected} and \ref{eq:noise}, which describe alignment and diffusion of the polarity direction.\\ 
\\Free parameters determining the initial conditions of the model are the number of particles $N$, the radius of the particles $\lambda$, the radius of the sphere $R$, the surface particle density $\rho = N/4\pi R^2$ and the coverage $c = N\pi\lambda^2 /4\pi R^2$. Based on the typical number of cells in the experimental data, we choose $N = 100$ and use the measured average surface density of the experimental data $\langle\rho_{\text{exp}}\rangle \approx 0.004 \ \mu\text{m}^{-2}$ to set the radius of the sphere $R \approx 43 \ \mu\text{m}$, which is close to the radii in the experiment. Furthermore, we choose $c = 1$ as we do not observe empty regions between cells. This also sets the radius of particles to $\lambda \approx 8.6 \ \mu \text{m}$. Using the experimental density in our simulation also allows us to rescale length using a length scale $l = 1 \ \mu \text{m}$. We set the time scale of our simulation to the time scale of self-propulsion $\tau = \frac{l}{v_0}$. Then, to solve equations \ref{eq:v_projected}-\ref{eq:noise}, we employ an Euler-Maruyama scheme in the following algorithm:
\begin{enumerate}
	\item Initialize the $N$ particles by distributing them equidistantly on the sphere \cite{sphere}.     
	\item Initialize random polarities and project them onto the surface of the sphere and normalize them afterwards. We find that in the parameter regimes of interest, our results are not sensitive to the initial configuration of the polarities.
	\item Propagate the polarities by equation \ref{eq:polarity_projected_rotation} and \ref{eq:polarity_projected}. This equation describes a rotation of the polarities in the tangential plane at the particle positions. As we use finite time steps, this changes the length of the polarities. Therefore, we normalize the polarities after we performed this step.
	\item Perform one Euler-step of equation \ref{eq:v_projected} which displaces all particles in a tangential direction which is calculated from the tangential projection of traction force and repulsion interaction. Note that repulsion forces are simply calculated in 3D Cartesian space. Because we use finite time steps, this will result in small particle displacements away from the sphere. We normalize position vectors afterwards and multiply them with the radius of the sphere so that particles stay effectively on the sphere.
	\item The displacement of particles also leads to a displacement of polarities. However, a new position defines a new tangential plane which means that old polarities are no longer tangential at the new position. Therefore, we project the polarities again on the sphere and normalize them which allows us to repeat step 3 - step 5 during our simulation.
\end{enumerate}

Throughout, we use $T = 4*10^{5}$ time steps and used a time step $\Delta t = 0.5\times10^{-2} \tau$. These parameters set the timescale of our simulation in relation the experiment. We use the observation that cell spheroids make between $1-2$ full revolutions within the experimental observation time of $23.33 \ \text{h}$ to determine $T$ after setting $\Delta t$ to a sufficiently small value. We choose $r_{\text{inter}} = 2.4\lambda$ and $\epsilon/\gamma = 0.8 \ \tau^{-1}$. This choice of the stiffness yields an approximately covered sphere while also avoiding jamming of particles.\\ 
\\For the phase diagram shown in main text Fig. 3c, we measure the noise amplitude $\sigma$ in units of $\tau^{-1}$ and vary it logarithmically in 20 steps in $[10^{-2}, 10^0]$. We measure the alignment strength $\beta$ in units of $\tau^{-1}$ and vary it logarithmically in 25 steps in $[10^{-2}, 10^{\frac{1}{2}}]$. For the phase diagram shown in main text Fig. 3c, we simulate $10$ realizations for each parameter combination. Near the transition between the collectively rotating regime and the quiescent regime, we additionally simulate $20$ realizations. We record the trajectories at a rate of $0.001$, yielding $400$ time points per trajectory which results in an effective time step in the sampled trajectories of $\Delta t = 5\tau$. We do not record any data during the first half of the simulation, which ensures that the system has reached steady state. This way we end up with $200$ time steps which yields an amount of simulation data that is comparable to the amount of experimental data.

%------------------------------------------------------------
\section{Supplementary model results}

%------------------------------------------------------------
\subsection{Robustness of velocity waves and parameter overview}

\begin{figure}[h]
	\centering
	\includegraphics[width=\textwidth]{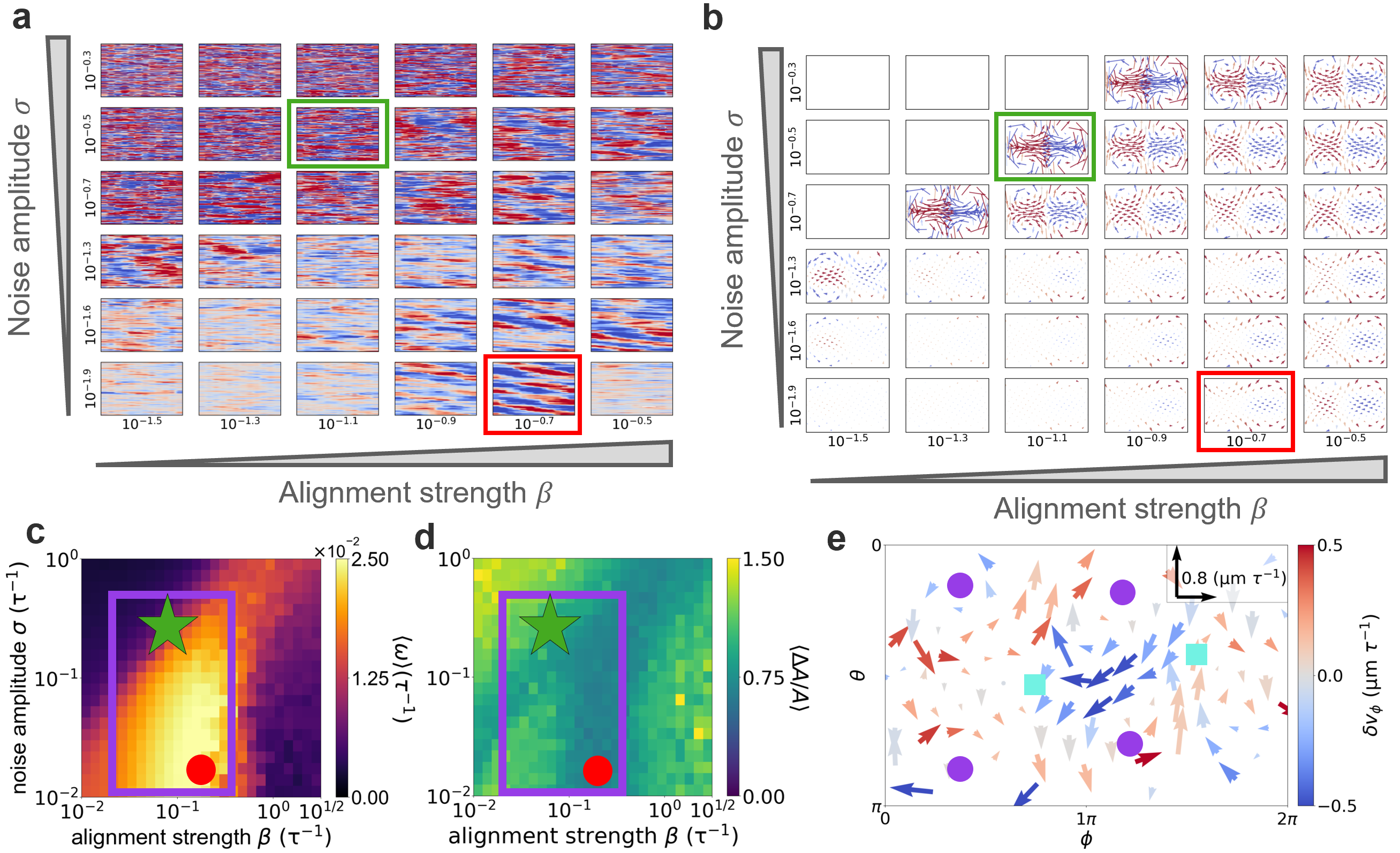}
	\caption{\textbf{Robustness of velocity wave.} \small \textbf{a)} Kymographs of the azimuthal equatorial velocity fluctuation profile $\delta v_{\phi}(\phi,t)$ for the parameter regime where we observe collective rotations in our model. The green box indicates the experimental parameter regime that quantitatively captures the experimental data. The red rectangle indicates a parameter combination within the low-noise parameter regime of our model. \textbf{b)} Tangential components of the average  velocity fluctuation field in the frame co-moving with the velocity wave for different parameter combinations. We restrict the noise-to-signal ratio of our fit to $\Delta A/A < 1.2$ and the average rotational order parameter $\langle\Omega\rangle > 0.6$. \textbf{c)} Phase diagram constructed from the average angular speed as discussed in the main text. The violet rectangle indicates the region for which we show the velocity wave in a and the average velocity fluctuation field in b. The green star indicates the point in the parameter space where the model captures the experimental data. The red point indicates the low-noise regime of our model. \textbf{d)} Noise-to-signal ratio for the parameter range we consider in our model. \textbf{e)} Snapshot of the tangential velocity fluctuation field represented in spherical coordinates for the experimental parameter regime (green star). The snapshot reveals the presence of two distinct regions in the equator where one region fluctuates in positive $\phi$-direction and the other in negative $\phi$-direction with saddle-point defects in between (turquoise squares). Above an below these regions, we observe vortices (violet circles) of polar motion.}
	\label{fig:robustness_wave}
\end{figure}

The velocity wave that we observe in our model is a robust feature in a wide range of parameters within the rotating regime. To show this, we consider the kymograph of the azimuthal equatorial velocity fluctuation profile $\delta v_{\phi}(\phi,t)$ for each parameter combination. This kymograph reveals a robust sinusoidal wave profile within the collectively rotating regime (Fig. \ref{fig:robustness_wave}a,c). We quantify the robustness by the noise-to-signal ratio of our fit defined by equation \ref{eq:noise_to_signal}. This ratio reveals a robustly significant sinusoidal wave profile ($\Delta A/A < 1.2$) within the rotating regime (Fig. \ref{fig:robustness_wave}d). Within this regime, we compute the average tangential velocity and velocity fluctuation fields following the procedure outlined in section \ref{sec:averaging}. We show the result of this procedure in Fig. \ref{fig:robustness_wave}b which reveals the robust presence of the pattern with four vortices within the rotating regime. Similar to the experiment, snapshots of the tangential velocity fluctuation field in spherical coordinates reveal that the average velocity fluctuation field are representative for individual time points (Fig. \ref{fig:robustness_wave}e). 

\subsection{Density modulations in the simulation}

To investigate possible density modulations in our model, we perform the same analysis (section \ref{sec:density}) for the simulation data. We find that our model is consistent with the experimental data: In high-repulsion regimes ($\epsilon = 5 \ \tau^{-1}$, $\beta = 0.079 \ \tau^{-1}$, $\sigma = 0.31 \ \tau^{-1}$), our model predicts no significant density modulation (Gray curve in Fig. \ref{fig:density}e). However, in this regime, effects from jamming and close packing of particles on a sphere become relevant as we noticed in a decreasing magnitude of velocity fluctuations. Therefore, to capture the magnitude of the velocity fluctuations in the experiment, we use weaker repulsion interactions in our standard parameter combination ($\epsilon = 0.8 \ \tau^{-1}$, $\beta = 0.079 \ \tau^{-1}$, $\sigma = 0.31 \ \tau^{-1}$). In this parameter combination, we find small density modulations ($\langle\delta\rho_{\text{norm}}(\phi,t)\rangle_{t,s} \approx 0.01$) in the frame of reference co-moving with the velocity wave, with a maximum ahead of the velocity wave minimum and behind the velocity wave maximum (Black curve in Fig. \ref{fig:density}e). This indicates that we operate in a regime of our model with small but finite compressibility. Since the density modulations are small compared to the average density and our model predicts insignificant density modulations in stiff regimes that also show velocity waves, we still conclude that our model approximates the incompressibility of the surface layer.

%------------------------------------------------------------
\subsection{Testing an alternative alignment interaction}

To test if the emergence of velocity waves in our model is specific to the way we implement active aligning motion in our model, we consider a second model for self-propelled particles. The model discussed in the main text and described in section \ref{sec:model} features polarity as an independent degree of freedom next to velocity. To test if the choice of the polarity as an independent degree of freedom is important for the emergence of velocity waves, we additionally consider a self-propelled particle model where polarity is now enslaved to the velocity. We again model cells as overdamped particles moving on a substrate with effective friction $\gamma$ \cite{Copenhageneaar8483}. The particles are self-propelled with speed $v_0$ into the direction of their polarity $\vec{p}_i$, which is now given in terms of the particle velocity. Furthermore, the polarity dynamics include dynamical Gaussian white noise $\vec{\eta}$ with amplitude $\sigma$. In this model, the particles interact via repulsion and alignment interactions, which in contrast to the model in the main text, now act on the velocity of the particles. The particles obey the following stochastic equations of motion:
\begin{equation}\label{eq:v_projected_velocity_model}
	\vec{v}_i(t+\Delta t) = \mathbf{P}_{T}\left(v_0\frac{\vec{p}_i(t)}{|\vec{p}_i(t)|} + \frac{1}{\gamma}\vec{F}_{\text{rep}}^{i}(r_{ij})\right)
\end{equation}
\begin{equation}\label{eq:repulsion_projected_velocity_model}
	\vec{F}^{i}_{rep}(r_{ij}) = -\varepsilon\sum_{r_{ij} < r_{\text{inter}}}(2\lambda - r_{ij})\hat{r}_{ij} 
\end{equation}
\begin{equation}\label{eq:polarity_velocity_model}
	\vec{p}_i(t) = \vec{v}_i(t) + \beta\frac{\vec{V}_i(t)}{|\vec{V}_i(t)|} + \vec{\eta}(t)
\end{equation}
\begin{equation}\label{eq:noise_velocity_model}
	\langle \eta_i(t)\eta_j(t')\rangle = \sigma^2 \delta(t-t')\delta_{ij} 
\end{equation} 
Again $\mathbf{P}_{T}$ is an operator that projects vectors onto the tangential plane of the particles. Note that we cut off the repulsion potential beyond $\lambda$, which implements pure repulsive interactions.\\
\\Alignment interactions are implemented by the term $\beta\frac{\vec{V}_i(t)}{|\vec{V}_i(t)|}$ in equation \ref{eq:polarity_velocity_model}, where $\vec{V}_i(t)$ is the average velocity of the $M$ particles in the vicinity of particle $i$ defined by $r_{ij} < r_{\text{inter}}$:
\begin{equation}
	\vec{V}_i(t) = \frac{1}{M}\sum_{r_{ij} < r_{\text{inter}}}\vec{v}_j(t)
\end{equation}
We initialize the particles and solve their equations of motion using the same scheme as for the main model explained in section \ref{sec:model}. We choose $\epsilon/\gamma = 0.01 \ \tau^{-1}$ and again $r_{\text{inter}} = 2.4\lambda$, where $\lambda$ is again the cut-off of the repulsive force between the particles.

\begin{figure}[h]
	\centering
	\includegraphics[width=\textwidth]{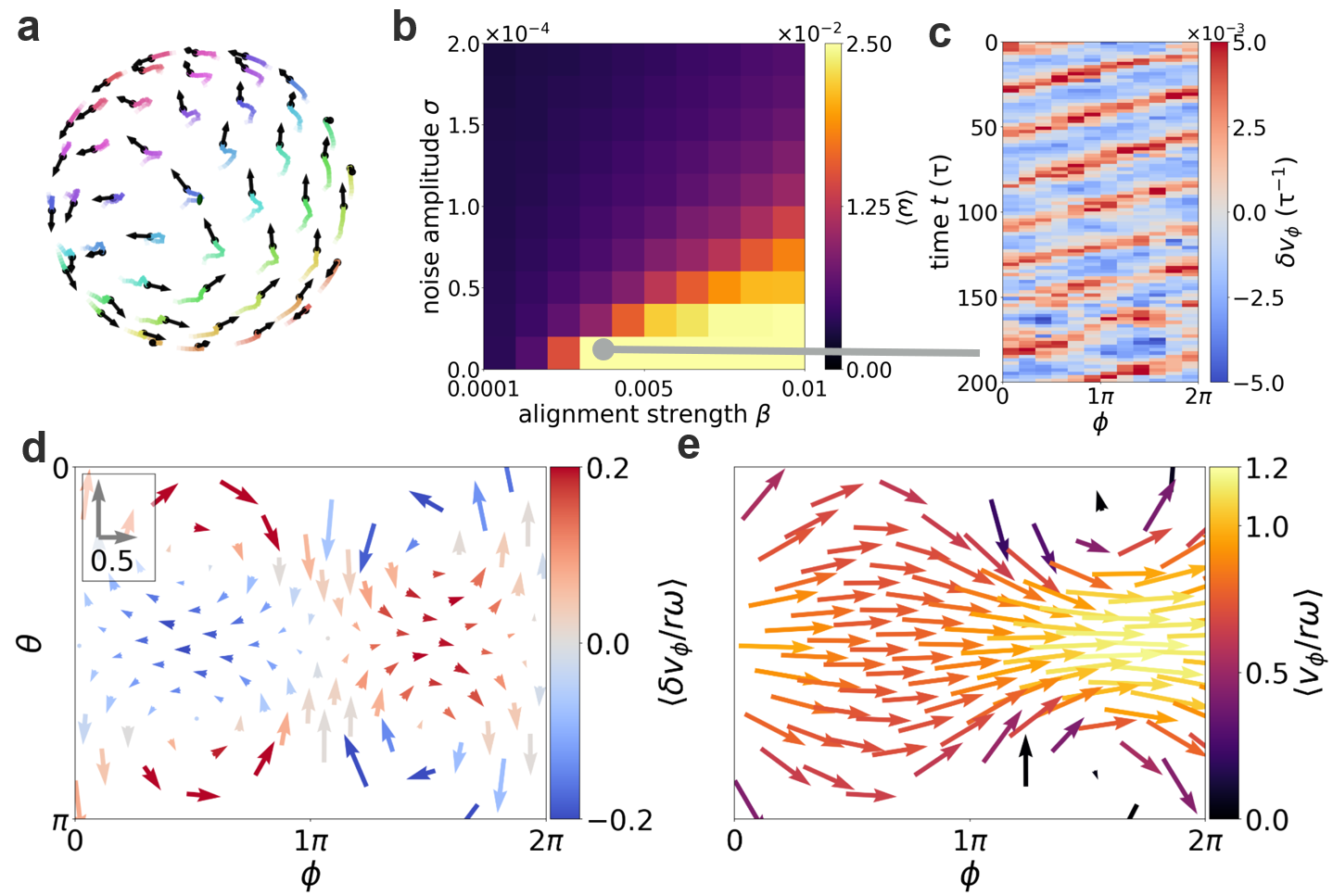}
	\caption{\textbf{Velocity waves from velocity alignment.} \small \textbf{a)} Snapshot of the rotational dynamics predicted by a model with alternative alignment interactions. \textbf{b)} Phase diagram of the alternative model constructed from the average angular speed $\langle\omega\rangle$. \textbf{c)} Kymograph of the azimuthal component of the velocity fluctuation field in the equator. Grey line shows the parameter combination, that we chose for this kymograph. ($\beta = 0.0045$, $\sigma = 0$) \textbf{d)},\textbf{e)} Tangential components of the average velocity fluctuation field (d) and the average total velocity field (e) in spherical coordinates as predicted by the model with alternative alignment interactions.}
	\label{fig:velocity_alignment}
\end{figure}
We again reveal a phase diagram for this model by computing the average rotational order and the average angular velocity of the particles (Fig. \ref{fig:velocity_alignment} b). This phase diagram resembles the one we obtain for the model discussed in the main text (Main text Fig. 3c). Importantly, this model also predicts propagating equatorial velocity waves (Fig. \ref{fig:velocity_alignment} c). These waves are robustly accompanied by a global pattern of migration characterized by four vortex defects and two saddle-point defects in the velocity fluctuation field as well as two vortex defects in the total velocity field (Fig. \ref{fig:velocity_alignment} d,e). These results show that the emergence of velocity waves is not specific to our choice of model, but only requires the generic features self-propulsion, alignment interactions, and a spherical geometry. 

\subsection{Analysis of rotational dynamics of the model}

\begin{figure}[h]
	\centering
	\includegraphics[width=\textwidth]{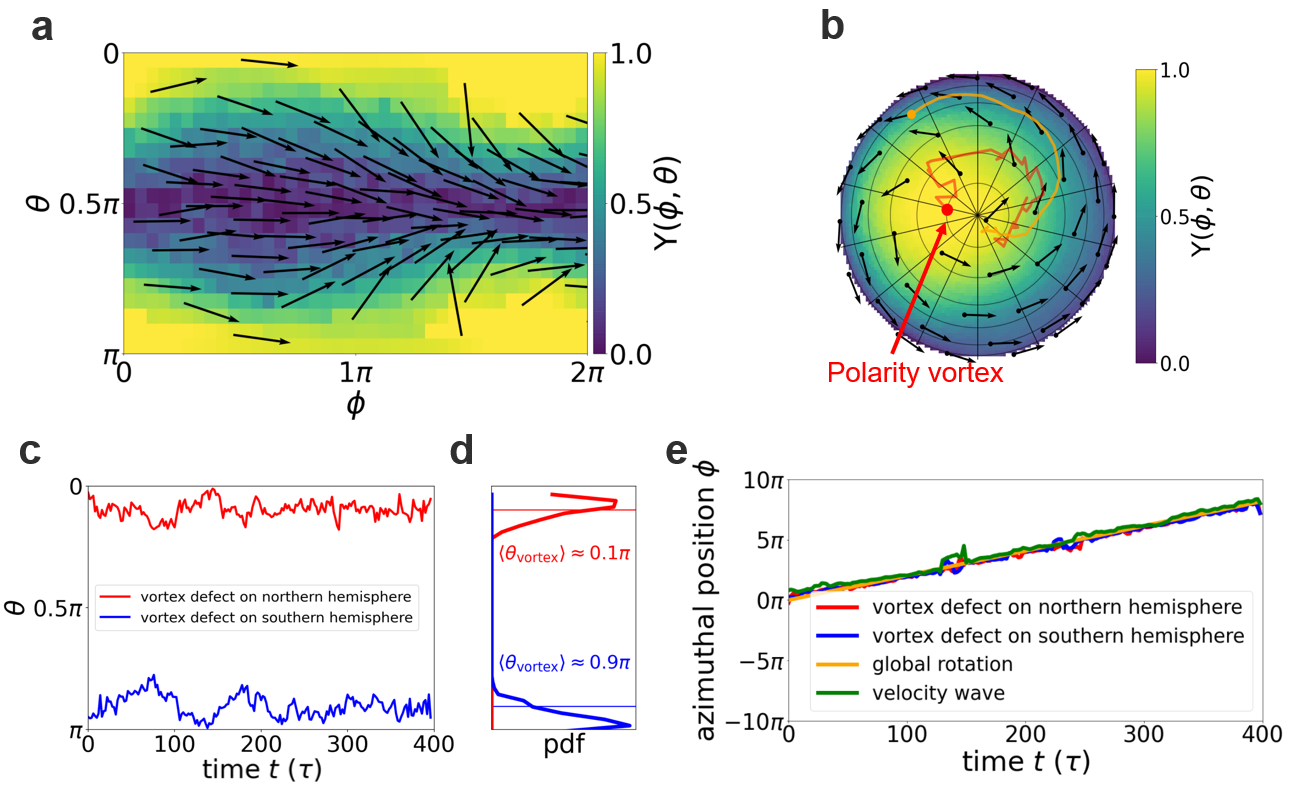}
	\caption{\textbf{Rotational dynamics of active particles.} \small \textbf{a)},\textbf{b)} Snapshot of the polarity field in our model in front of its vorticity measure $|\Upsilon(\phi,\theta)|$. a) shows a spherical projection of both quantities, b) shows a planar projection of the northern hemisphere. \textbf{c)} Trajectory of the polar angles $\theta_{\text{northern\ vortex}}$ and $\theta_{\text{southern\ vortex}}$ of the two polarity vortex defects. \textbf{d)} Probability distribution of the polar angles of the polarity defects. \textbf{e)} Trajectory of the azimuthal angles of the two polarity vortex defects, the global rotation, and the equatorial velocity wave.}
	\label{fig:propagating}
\end{figure}

To characterize the rotational dynamics of our main model, we investigate the vortex defects in the polarity field of our simulation at a low-noise parameter combination of our model ($\beta = 0.2 \ \tau^{-1}$, $\sigma = 0.012 \ \tau^{-1}$). We find the position of these defects by computing the vorticity measure $\Upsilon(\phi,\theta)$ of the polarity field in the way we defined for the tangential velocity fluctuation field (section \ref{sec:vorticity}). We infer the maxima of $|\Upsilon(\phi,\theta)|$, which correspond to the positions of vortices on the northern and southern hemisphere in spherical coordinates (Fig. \ref{fig:propagating}a,b). We find that these polarity vortices are displaced away from the poles of the average rotation (Fig. \ref{fig:propagating}c,d). The vortex displacements are given in spherical coordinates by
\begin{equation}
	\theta_{\text{northern\ vortex}} \approx 0.1\pi, \ 1 - \theta_{\text{southern\ vortex}}  \approx 0.1\pi
\end{equation}
Here, $(\phi_{\text{northern\ vortex}},\theta_{\text{northern\ vortex}})$ is the position of the vortex defect on the northern hemisphere in spherical coordinates. Furthermore, we find that these polarity defects propagate around the axis of rotation (Fig. \ref{fig:propagating}b,e). 

Interestingly, the azimuthal angle of the displaced vortices coincides with the azimuthal angle of the velocity wave. Furthermore, the speed with which the polarity vortices are propagating is remarkably similar to the propagation speed of the velocity wave that also propagates around the axis of rotation (Fig. \ref{fig:propagating}e). Note that in contrast to the experimental data, in the low-noise regime of our simulation does not show significant fluctuations of the axis of rotation. Therefore, we can assume that the axis of rotation is approximately fixed in space $\hat{\omega}(t) \approx \hat{\omega}(t + \Delta t)$ and we can use the inferred trajectories of the vortices and the velocity wave with respect to the axis of rotation to characterize the 3D COM trajectory of the vortices and velocity wave. The coupling between the velocity wave and the vortices reveals that the velocity wave in the equator is robustly coupled to a pattern in the polarity field that features polarization over the poles of the global rotation and a displaced polarity vortex defect. This shows that the velocity wave is not a random large scale velocity fluctuation in the model, but driven by active self propulsion towards and over the poles.\\
\\Finally, our analysis reveals long periods in time with no apparent differences between the speed of the velocity wave, vortex motion, and global rotation in the low-noise regime of our model (Fig. \ref{fig:propagating}e), which is consistent with the experimental results (Fig \ref{fig:wave_propagation}).
%\\To further characterize the rotational dynamics we observe here, we analyze single-particle trajectories. In particular, we focus on particles that are initially part of the polarity vortex (Fig. \ref{fig:propagating}b). We find that, on longer time scales, these particles will no longer be a part of the vortex but move towards the equator. This suggests that particles do not rotate around the vortex defect. Instead, particles moving over the poles of the global rotation, move past the vortex, being only part of it for short periods of time. The direction of the motion over the poles is not static, but rotates around the axis of rotation, which consequently also displaces polarity vortices around the axis of rotation. These results reveal the unique rotational dynamics of active particles being constrained on a sphere. In contrast to rigid-body rotation, active particles do not perform azimuthal motion along circles around a common axis of rotation. Instead they exhibit pronounced non-azimuthal motion over the poles of the global rotation, which is characterized by displaced polarity vortices and converging flow into the equator, where polar fluxes are converted in equatorial velocity waves (Fig. \ref{fig:density}f).

\subsection{Disentangling the role of topology and curvature}

To disentangle the roles of the closed topology and geometric curvature in the emergence of velocity waves, we perturb the spherical geometry of our model. 

\subsubsection{Truncated sphere}
First, we consider a truncated sphere: we exclude two diametrically-opposed spherical caps defined by a polar angle of $\Theta = \pi/6$ (Fig. \ref{fig:geometry}b). We implement these regions by defining soft circular boundaries consisting in a linear repulsive force. The two boundaries confine the particles on a ring with positive curvature which is an open manifold and thus topologically distinct from the closed spherical geometry. We adjust the number of particles to keep the density, the size of the particles, and the radius of the sphere constant. With a radius of the sphere $R = 43 \mu \mathrm{m}$, the excluded region defines a circle that spans five times the radius of the particles. Therefore, two particles on two diametrically-opposed points of one excluded region do not interact with each other (radius of interaction $r_{\text{inter}} = 2.4\lambda $).\\ 
\\We consider a parameter combination that predicts almost deterministic rotations on the sphere ($\beta = 0.1 \ \tau^{-1}$, $\epsilon = 0.8 \ \tau^{-1}$, $\sigma = 0.005 \ \tau^{-1}$) (Fig. \ref{fig:geometry}a,e, Supplementary Movie 5). Using this combination on the truncated sphere, we test whether the change of geometry leads to a loss of the velocity wave. Performing numerical simulations reveals that particles perform global rotations along the boundaries around the two excluded regions (Fig. \ref{fig:geometry}b). We find that the velocity wave still emerges (Fig. \ref{fig:geometry}f), together with global patterns of velocity fluctuations (Fig. \ref{fig:geometry}i,j). Furthermore, we consider a parameter combination that is closer to the experimental parameter regime of the sphere ($\beta = 0.15 \ \tau^{-1}$, $\epsilon = 0.8 \ \tau^{-1}$, $\sigma = 0.15 \ \tau^{-1}$), which we used to reproduce the experimental data. Here, we also observe a velocity wave accompanied by a global pattern of motion (Fig. \ref{fig:geometry}g,k). The average pattern in the experimental parameter regime is shown in main text Figure 3o. In conclusion, our results suggest that the velocity wave still emerges on truncated spheres that are topologically distinct from spheres. 

\subsubsection{Active particles on cylinders}

To further disentangle the role of topology and curvature, we consider a cylinder, i.e. a geometry with the same topology as the truncated sphere but with zero Gaussian curvature. We implement this model by constraining the dynamics of particles on the surface of a cylinder. Specifically, we use equations \ref{eq:v_projected} - \ref{eq:noise}, but implement the cylindrical geometry by using a different normal vector $\hat{n}$ of the underlying surface. For the sphere, we used the normalized position vector of a cell $\hat{r}_i$ as normal vector. Projection of the forces on the tangential plane of a particle was done using this vector (equation \ref{eq:v_projected}). Also rotation in the tangential plane of polarity vectors due to noise and alignment interactions was done using this vector (equation \ref{eq:polarity_projected_rotation} and \ref{eq:polarity_projected}). For the cylinder, we replace this vector by the normal vector of the cylindrical surface $\hat{n}_i = (\cos(\phi_i), \sin(\phi_i), 0)$ at the position of the i-th cell. Finally, we introduce two soft boundaries perpendicular to the z-axis of the cylinder and separated by a height $h$. We choose $N = 150$ particles, and use the same experimental surface density of particles as we did with the sphere. 
\begin{figure}[h]
	\centering
	\includegraphics[width=\textwidth]{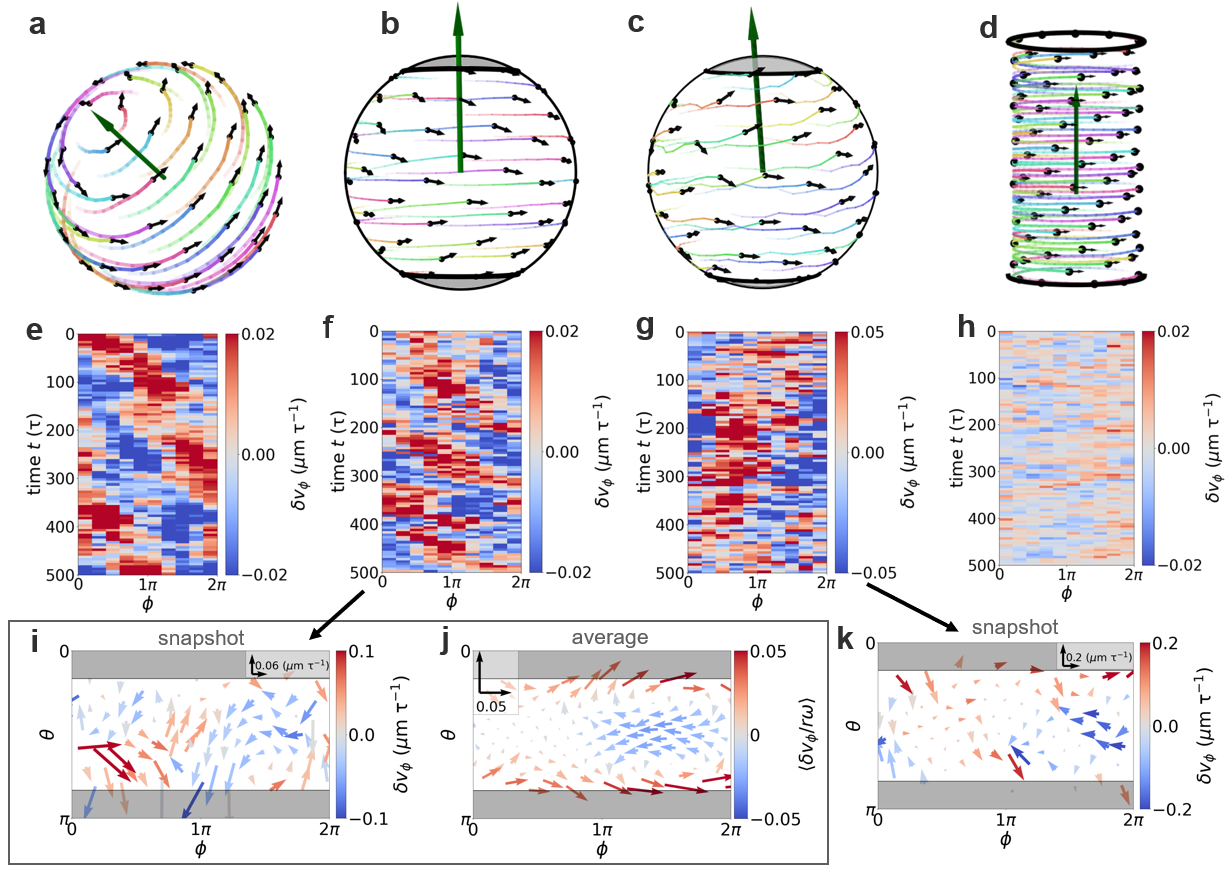}
	\caption{\textbf{Active particles on varying geometries.} \small \textbf{a)}-\textbf{d)} Snapshots of the dynamics of active particles on a sphere (a), on a truncated sphere with deterministic rotations (b), on a truncated sphere with noisy rotations capturing closer to the experimental parameter regime (c), and on a cylinder (d). For a,b, and d we use the same parameter combination ($\beta = 0.1$, $\epsilon = 0.8$, $\sigma = 0.005$). \textbf{e)} Kymograph of equatorial velocity fluctuations $\delta v_{\phi}(\phi,t) - \langle \delta v_{\phi}(\phi,t) \rangle_{\phi}$ for the sphere (e), on a truncated sphere with deterministic rotations (f), on a truncated sphere with noisy rotations capturing closer to the experimental parameter regime (g), and on a cylinder (h). Note that for the cylinder, we use the same procedure as for the kymograph in spherical geometries with the only change of constraining the z-direction between $h/3$ and $-h/3$ instead of using the polar angle $\theta$ to focus on the equatorial region. We subtract $\langle \delta v_{\phi}(\phi,t) \rangle_{\phi}$ from the kymograph $\delta v_{\phi}(\phi,t)$ because we sometimes saw an offset of the velocity wave. (See equatorial region in panel j) \textbf{i)} Snapshot of the tangential velocity fluctuation field in spherical coordinates for the truncated sphere with deterministic rotations. \textbf{j)} Average tangential velocity fluctuation in the frame co-moving with the velocity wave in truncated spheres in the parameter regime of deterministic rotations. \textbf{k)} Snapshot of the tangential velocity fluctuation field in spherical coordinates for the truncated sphere with more stochastic rotations. The average of these patterns is shown in main text Fig. 3o.}
	\label{fig:geometry}
\end{figure}
This means that the radius of particles is set to $8.6 \ \mu\text{m}$ at a coverage $c \approx 0.9$. We furthermore set the radius of the cylinder to $R \approx 36 \ \mu\text{m}$, which is comparable to the size of the spheres that we studied. Together with the number of particles and the coverage this yields a cylinder height of $h \approx 172 \ \mu\text{m}$.\\
\\Changing the geometry from a sphere to a cylinder requires an additional adjustment of the alignment interactions of the model. Note that we implement alignment interactions on the sphere by finding the angle between two polarity vectors and then rotating the polarity of a cell in the tangential plane to decrease this angle. Importantly, the angle is computed as the projection of the cross product of the two 3D polarity vectors on the surface normal (equation \ref{eq:polarity_projected_rotation} and \ref{eq:polarity_projected}). Applying the same alignment interaction on the cylinder leads to very stable states where all polarities are aligned along the z-axis of the cylinder, which prohibits collective rotations. 
%This is because no matter what the configuration of 3D polarity vectors on a cylinder is, there is always a non-zero angle between two 3D polarity vectors causing rotation of these vectors towards the z-direction making them completely parallel. 
For a sphere, such a special state where all 3D polarities can be completely aligned does not exist. Therefore, we adjust the alignment interaction for the cylinder: instead of finding the angle between the 3D polarity vectors, we parallel-transport the polarity vectors onto each other to find an angle between them in one local tangent plane. This implementation thus considers vectors to be parallel when they have the same components in cylindrical coordinates, which no longer leads to a favored direction, and thus allows persistent rotations around the z-axis. We also performed tests to show that this alternative alignment interaction does not affect the collective dynamics of the particles on a sphere. In these test, we found no differences in the rotational behavior as well as in the properties of the velocity wave.\\
\\We initialize the particles on the cylinder with purely azimuthal polarities and solve the model numerically with $T = 4*10^{5}$ time steps and a time step $\Delta t = 0.5\times10^{-2}\ \tau$. We record trajectories at a rate of $0.001$. We choose $\sigma = 0.005 \ \tau^{-1}$, $\epsilon = 0.8 \ \tau^{-1}$ and $\beta = 0.1 \ \tau^{-1}$ and repeat the simulation $5$ times. This is the same parameter combination that we used for both the sphere (Fig. \ref{fig:geometry}a) and the truncated sphere (Fig. \ref{fig:geometry}b). We analyze the emerging global rotations on the cylinder (Fig. \ref{fig:geometry}d, Supplementary Movie 6) and look for velocity waves in the azimuthal component of the velocity fluctuation fields as we did with the sphere. First, velocity fluctuations are smaller compared to the complete and the truncated spheres, indicating that the global rotations are barely modulated on a cylinder (Fig. \ref{fig:geometry}h). Furthermore, we are not able to detect any large-scale velocity fluctuations modulating the global rotation while persistently propagating in the direction of the global rotation. This suggests that particles on a cylinder perform pure azimuthal motion along parallel circular trajectories. When comparing to the truncated sphere that has the same topology as a cylinder but features non-zero Gaussian curvature, the absence of velocity waves on cylinders indicates that Gaussian curvature is required for the emergence of velocity waves.

\newpage

\bibliographystyle{biolett}
\bibliography{Verzeichnis}

\end{document}